\documentclass[a4paper,12pt]{article}
\usepackage{jheppub}
\usepackage[dvipsnames]{xcolor}
\usepackage{tikz}
\usepackage{pgfplots}
\usepackage[T1]{fontenc}
\usepackage[]{slashed}
\usepackage[]{bm}     
\usepackage{physics}
\usepackage{dsfont}

\usepackage[force]{feynmp-auto}

\usepackage{hyperref}
\hypersetup{colorlinks=true,linkcolor=magenta,anchorcolor=green,citecolor=cyan,filecolor=black,menucolor=black,urlcolor=brown}
\makeatletter

\usepackage{lipsum}

\DeclareMathOperator\arcsinh{arcsinh}
\DeclareMathOperator\arccosh{arccosh}

\DeclareMathOperator\Li{Li}

\usepackage{graphicx}
\usepackage{epstopdf}
\usepackage{bbm,amsmath,graphicx,amssymb,amsfonts,amsthm}

\def\sq[#1,#2]{\left[#1\,#2\right]}
\def\an[#1,#2]{\left\langle#1\,#2\right\rangle}

\def\an[#1,#2]{\left\langle#1\,#2\right\rangle}
\def\aq[#1,#2,#3]{\left\langle#1|#2|#3\right]}
\def\qa[#1,#2,#3]{\left[#1|#2|#3\right\rangle}
\def\sq[#1,#2]{\left[#1\,#2\right]}
\def\spa#1.#2{\left\langle#1\,#2\right\rangle}
\def\spab[#1,#2,#3]{\left\langle#1|#2|#3\right]}
\def\spba[#1,#2,#3]{\left[#1|#2|#3\right\rangle}
\def\spb#1.#2{\left[#1\,#2\right]}

\def\Ttrma(#1,#2,#3,#4){{\rm tr}_{-}[\slash \!\!\!\;\!\! #1\slash  \!\!\!\;\!\! #2 \slash  \!\!\!\;\!\!#3\slash  \!\!\!\;\!\!#4]}
\def\Ttrmb(#1,#2,#3,#4,#5,#6){{\rm tr}_{-}[\slash \!\!\!\;\!\! #1\slash  \!\!\!\;\!\! #2 \slash  \!\!\!\;\!\!#3\slash  \!\!\!\;\!\!#4\slash  \!\!\!\;\!\!#5\slash  \!\!\!\;\!\!#6]}
\def\Ttrmc(#1,#2,#3,#4,#5,#6,#7,#8){{\rm tr}_{-}[\slash \!\!\!\;\!\! #1\slash  \!\!\!\;\!\! #2 \slash  \!\!\!\;\!\!#3\slash  \!\!\!\;\!\!#4\slash  
\!\!\!\;\!\!#5\slash  \!\!\!\;\!\!#6\slash  \!\!\!\;\!\!#7\slash  \!\!\!\;\!\!#8]}
\def\Dp(#1,#2){(#1\cdot #2)}

\def\triangleboxleft{\scalebox{.9}{$\triangleleft$}\kern-.1em\Box}
\def\triangleboxright{\Box\kern-.1em\scalebox{.9}{$\triangleright$}}
\def\dBox{\Box\kern-.1em\Box}
\def\dNPBoxs{\scalebox{.9}{$\bowtie$}\kern-.1em\Box}
\def\dNPBoxu{\Box\kern-.1em\scalebox{.9}{$\bowtie$}}

\def\beq{\begin{equation}}
\def\eeq{\end{equation}}
\def\bes{\begin{split}}
\def\ees{\end{split}}
\def\beqa{\begin{eqnarray}}
\def\eeqa{\end{eqnarray}}

\def\eeqa{\end{eqnarray}}
\def\ek[#1,#2]{(\varepsilon_{#1}\cdot k_{#2})}
\def\e[#1,#2]{(\varepsilon_{#1}\cdot \varepsilon_{#2})}
\def\s(#1,#2){{(\ell_#1\cdot\ell_#2)}}

\def\e{\epsilon}

\pgfdeclarelayer{bg}    
\pgfsetlayers{bg,main}  

\definecolor{Mathematica}{HTML}{ed192d}

\usepackage{float}

\usetikzlibrary{shapes.misc}
\usetikzlibrary{arrows}
\usetikzlibrary{decorations.markings}
\usetikzlibrary{positioning,fit}
\usetikzlibrary{patterns}

\tikzset{cross/.style={cross out, draw=black, minimum size=2*(#1-\pgflinewidth), inner sep=0pt, outer sep=0pt},
	cross/.default={2pt}}

 \preprint{\vbox{\hbox{\hphantom{XXXX}IPhT-t21/028}\hbox{\hphantom{X}CERN-TH-2021-073}}}
      
\title{The Amplitude for Classical Gravitational Scattering at Third Post-Minkowskian Order}

\author[a]{N. Emil J. Bjerrum-Bohr}
\author[a,d]{\!\!, Poul H. Damgaard}
\author[a]{\!\!, Ludovic Plant\'e}
\author[b,c,d]{\!\!, Pierre Vanhove}

\affiliation[a]{Niels Bohr International Academy, Niels Bohr Institute, University of Copenhagen, Blegdamsvej 17, DK-2100 Copenhagen, Denmark}
\affiliation[b]{Institut de Physique Theorique, Universit\'e Paris-Saclay,
CEA, CNRS, F-91191 Gif-sur-Yvette Cedex, France}
\affiliation[c]{National Research University Higher School of
  Economics, Russian Federation}
\affiliation[d]{Theoretical Physics Department, CERN, 1211 Geneva 23, Switzerland}

\abstract{We compute the scattering amplitude for classical black-hole scattering to third order in the Post-Minkowskian expansion, 
keeping all terms needed to derive the scattering angle to that order from the eikonal formalism. Our results
confirm a conjectured relation between  the real and imaginary parts of the amplitude by Di Vecchia, Heissenberg,
Russo, and Veneziano, and are in agreement with a recent computation by Damour based on radiation reaction
in general relativity.}

\begin{document} 
\maketitle
\flushbottom
\newpage
\section{Introduction}\label{sec:intro}

The Post-Minkowskian expansion of Einstein gravity is arguably one of the most fertile areas in which to apply modern methods of amplitude
calculations~\cite{Damour:2016gwp,Damour:2017zjx,Bjerrum-Bohr:2018xdl,Cheung:2018wkq}.  Being based solely on an expansion in Newton's
constant $G_N$, results are valid at all velocities and are thus effectively re-summing an infinite number of terms of the perhaps more familiar
Post-Newtonian expansion. Progress has been very rapid and first results for the conservative part of the interaction Hamiltonian
are now known to both third and fourth order in the Post-Minkowskian expansion~\cite{Bern:2019nnu,Bern:2019crd,Cheung:2020gyp,Bern:2021dqo}.
These computations are firmly rooted in established quantum field theoretic methods, be they phrased in terms of effective field theory matching
\cite{Cheung:2018wkq} or, equivalently, in terms of the standard definition of a two-body interaction potential from the
iterated solution of the Lippmann-Schwinger equation, which quantify exactly what is commonly known as Born subtractions~\cite{Cristofoli:2019neg}.
The conservative part of the scattering to third Post-Minkowskian has also been confirmed in the framework of the world-line approach~\cite{Kalin:2020fhe}. 
A remarkable relationship between the relativistic two-body kinematics and the iterated solution to the
Lippmann-Schwinger equation leads to explicit formulas for the scattering angle in terms of the coefficients of the relativistic two-body potential
\cite{Kalin:2019rwq,Bjerrum-Bohr:2019kec,Kalin:2019inp}, in a sense thereby solving the problem of kinematics of the relativistic
two-body problem in general relativity by means of quantum field theory and amplitude techniques.

The puzzling aspect of an ill-defined high-energy limit of the conservative part of the two-body interaction Hamiltonian at third Post-Minkowskian
order~\cite{Bern:2019crd} has recently led to surprising new insight into the relationship between the two-to-two gravitational scattering amplitude 
and general relativity~\cite{DiVecchia:2020ymx,Damour:2020tta,DiVecchia:2021ndb,DiVecchia:2021bdo}. From the amplitude point of view, refs.~\cite{DiVecchia:2020ymx,DiVecchia:2021ndb,DiVecchia:2021bdo} have shown how certain classical terms of the two-loop scattering, that are not captured
by a limitation to the so-called potential region of the loop integrals, restore a well-behaved high-energy behavior. This picture has been understood
directly from general relativity in terms of gravitational radiation reaction~\cite{Damour:2020tta}. Indeed, in the low-energy limit the new terms give rise to half-integer powers
in the Post-Newtonian expansion and thus belong to effects described by radiation. By using the formalism of ref.~\cite{Kosower:2018adc} a direct
computation of the radiated momentum to that order leads to the same result for the scattering angle~\cite{Herrmann:2021lqe,Herrmann:2021tct}
(see also~\cite{Parra-Martinez:2020dzs}).

A natural language for the calculation of the scattering angle of two black holes in general relativity is the gravitational eikonal. It relies on the exponentiation
of appropriate terms of the $S$-matrix in the small-angle limit of the involved semi-classical field theory amplitudes. Exponentiation has been proven to all orders
for Einstein gravity
at both leading Post-Minkowskian counting~\cite{Kabat:1992tb} and next-to-leading order for equal masses~\cite{Akhoury:2013yua} (generalized to different
masses in ref.~\cite{Bjerrum-Bohr:2018xdl}). This powerful framework
has led to the famous prediction for high-energy gravitational scattering
by Amati, Ciafaloni, and Veneziano~\cite{Amati:1990xe} and has recently been explored and extended in numerous directions 
\cite{Collado:2018isu,Ciafaloni:2018uwe,KoemansCollado:2019ggb,DiVecchia:2019myk,DiVecchia:2019kta,Bern:2020gjj,Cristofoli:2020uzm,DiVecchia:2020ymx,
DiVecchia:2021ndb,DiVecchia:2021bdo}. Although the eikonal formalism is used to derive the {\em classical} scattering angle, an interesting feature is that also
an in principle infinite number of super-classical terms (corresponding to inverse powers of $\hbar$) must be computed as well in order to confirm the exponentiation
of the amplitude in impact parameter space, a phenomenon that ultimately must follow from unitarity alone~\cite{Cristofoli:2020uzm}. Moreover, an intricate interplay 
between classical and quantum pieces of the amplitude conspire to provide the correct classical scattering angle at any given order in the Post-Minkowskian expansion. 

Crucial to the argument of refs.~\cite{DiVecchia:2020ymx,DiVecchia:2021ndb,DiVecchia:2021bdo} is a remarkably simple relation between the divergent
part of the imaginary part of the amplitude and the finite real part
of the radiation-reaction contribution. In maximal supergravity, this relation has recently 
been explicitly confirmed from the two-to-two scattering amplitude
\cite{Bjerrum-Bohr:2021vuf} as well as from the radiated momentum calculation~\cite{Herrmann:2021tct}. Extended to general relativity, the entirely different
approach of ref.~\cite{Damour:2020tta} should leave little doubt that this relation holds in Einstein gravity as well. Nevertheless, an explicit confirmation from
a full amplitude calculation seems needed at this stage. The purpose of the present paper is to fill this gap and, hopefully, provide some further insight into
the result. We shall employ the method recently described in ref.~\cite{Bjerrum-Bohr:2021vuf}. The idea is to organize the integrand of loop
amplitudes in subsets that naturally, due to the $i\varepsilon$-prescription of the Feynman propagator, combine into delta functions over momenta
in a manner reminiscent of eikonal calculations~\cite{Kabat:1992tb} and which have earlier also been used to simplify the evaluation of some of the involved loop
integrals~\cite{Bern:2019crd,Parra-Martinez:2020dzs}. Imposing these delta functions lowers the dimensionality of the integrals
in a covariant manner. A second advantage of the method of ref.~\cite{Bjerrum-Bohr:2021vuf} is that it significantly reduces the number of master integrals
that need to be known. This was obvious in the case of maximal
supergravity and, as we shall demonstrate in this paper, this holds in
Einstein gravity as well. No new master integrals are needed as compared to the supergravity case~\cite{Bjerrum-Bohr:2021vuf}.
In the end, we believe that we here for the first time compute all terms of the gravitational scattering amplitude of massive objects that are needed to obtain the classical
scattering angle to third Post-Minkowskian order. Simultaneously, we hope that this can help paving the way for more efficient evaluations at even higher orders. We also explicitly
confirm the relation between the divergent imaginary part and the real part of the radiation-reaction piece that was put forward in 
refs.~\cite{DiVecchia:2020ymx,DiVecchia:2021ndb,DiVecchia:2021bdo}.
\vfill\newpage
\section{Einstein gravity at two-loop order}\label{sec:3PMlevel}

Our starting point is the Einstein-Hilbert Lagrangian minimally coupled to two massive scalar fields:
\begin{equation}
{\cal L}_{EH} = \int d^4 x \sqrt{-g} \Bigg[\frac{R}{16 \pi G_N}  + \frac12 g^{\mu\nu} ( \partial_\mu \phi_1\partial_\nu \phi_1 + \partial_\mu \phi_2\partial_\nu \phi_2)- {m_1^2\over2} \phi_1^2 - {m_2^2\over2} \phi_2^2\Bigg]\,.
\end{equation} 
Here, $R$ defines the Ricci 
scalar and $g$ is the determinant of the metric: $g_{\mu\nu}(x)\equiv \eta_{\mu\nu} + \sqrt{32 \pi G_N}h_{\mu\nu}(x)$ expanded around a Minkowski background, ${\rm diag}\,\eta_{\mu\nu}\equiv(1,-1,-1,-1)$. 

$$
    \begin{fmffile}{kinvar}
    \begin{fmfgraph*}(100,100)
\fmfleftn{i}{2}
\fmfrightn{o}{2}
\fmfv{decor.shape=oval, decor.filled=shaded, decor.size=(.5w)}{v1}
\fmfv{decor.shape=circle,decor.filled=gray25,decor.size=(.5w)}{v1}
\fmf{fermion,label=$p_1$,label.side=left}{i1,v1}
\fmf{fermion,label=$p_1'$,label.side=left}{v1,i2}
\fmf{fermion,label=$p_2'$,label.side=right}{v1,o2}
\fmf{fermion,label=$p_2$,label.side=right}{o1,v1}
\end{fmfgraph*}
\end{fmffile}
$$

We consider here the  two-to-two amplitude with $p_1$ and
$p_2$ denoting incoming momenta and ${p_1'}$ and ${p_2'}$ outgoing momenta such that
$p_1^2 = {p_1'}^2 = m_1^2$ and $p_2^2 =
{p_2'}^2 = m_2^2$.

We work in the center of mass frame and our  conventions for the kinematical invariants are:
\begin{equation}\label{e:sdef} s\equiv(p_1+p_2)^2 = ({p_1'}+{p_2'})^2 
= m_1^2+m_2^2+2m_1 m_2 \sigma,\quad \sigma \equiv \frac{p_1 \cdot p_2}{m_1 m_2}\,, \end{equation} 
\begin{equation}\label{e:tdef}
t \equiv (p_1-{p_1'})^2 = ({p_2'}-p_2)^2\equiv q^2=-\vec{q}^{\,2}\,,\end{equation}
and 
\begin{equation}\label{e:udef}
u \equiv (p_1-{p_2'})^2 =  ({p_1'}-p_2)^2\,,
\end{equation}
so that, as usual, $s$ gives the center of mass energy, $E_{CM}^2$, and $t$ is
the transferred momentum. The classical limit is extracted by sending $\hbar\to0$ and
keeping $\underline q= q/\hbar$ fixed.

The two-loop amplitude we evaluate below is composed  by a piece
coming from the  three-particle cut with only gravitons propagating across the cut
\begin{equation}
\hspace{-1.5cm}  {\mathcal M}^2_{\rm 3-cut}(\sigma,q^2)=
\hspace{.4cm}  \begin{gathered}
\begin{fmffile}{3cut}
 \begin{fmfgraph*}(100,100)
\fmfstraight
\fmfleftn{i}{2}
\fmfrightn{o}{2}
\fmftop{t}
\fmfbottom{b}
\fmfrpolyn{smooth,filled=30,label=\textrm{tree}}{el}{8}
\fmfrpolyn{smooth,filled=30,label=\textrm{tree}}{er}{8}
\fmf{fermion,label=$p_1$,label.side=left,tension=2}{i1,el1}
\fmf{fermion,label=$p_1'$,label.side=left,tension=2}{el3,i2}
\fmf{fermion,label=$p_2'$,tension=2}{er5,o2}
\fmf{fermion,label=$p_2$,label.side=right,tension=2}{o1,er7}
\fmf{dbl_wiggly,tension=.1}{el5,er3}
\fmf{dbl_wiggly,tension=.1}{el6,er2}
\fmf{dbl_wiggly,tension=.1}{el7,er1}
\fmf{dashes,for=red}{b,t}
\end{fmfgraph*}
\end{fmffile}
  \end{gathered}
\end{equation}
and another piece from the
self-energy and vertex correction contributions given in section~\ref{sec:selfenergy}
\begin{equation}
  \mathcal M_2(\sigma,q^2)=   \mathcal M_2^{\rm
  3-cut}(\sigma,q^2)+\mathcal M_2^{\rm self-energy}(\sigma,q^2)\,.
\end{equation}
The three particle-cut is defined in $D=4-2\epsilon$ dimensions as
\begin{multline}\label{e:3cut}
\mathcal M_{2}^{\rm 3-cut}(\sigma,q^2)=\int \frac{d^D l_1
  d^D
  l_2d^Dl_3}{(2\pi)^{3D}} (2\pi)^D\delta^{(D)}(l_1+l_2+l_3+q){i^3\over
l_1^2l_2^2l_3^2}\cr\times{1\over3!}\sum_{\textrm{Perm}(l_1,l_2,l_3)\atop
  \lambda_1=\pm,\lambda_2=\pm,\lambda_3=\pm} \mathcal M_0(p_1,p_1',l_1^{\lambda_1},l_2^{\lambda_2},l_3^{\lambda_3})(\mathcal M_0(p_2,p_2',-l_1^{\lambda_1},-l_2^{\lambda_2},-l_3^{\lambda_3}))^{*},
\end{multline}
which involves the five points tree-level amplitudes given in the appendix~\ref{sec:fiveamplitude}.
The sum is over the helicity configuration $\lambda_i$ of the
gravitons across the cut. We only  keep the cut-constructible part of
the two-loop amplitude.

The three-particle cut gives both the conservative part of the
classical potential plus some of the radiation-reaction pieces. A few self-energy and vertex corrrection diagrams are needed to get
the full radiation-reaction term. 
Based on the $\hbar$ counting in eq.~(3.6)
of~\cite{Bjerrum-Bohr:2021vuf}  it is clear the one needs a least two
massive propagators to get a classical contribution to the
amplitude. By inspection of the possible topology of the Feynman
graphs we conclude that the three-particle cut contribution, the
self-energy and vertex corrrection diagrams are the only one
contributing to this order.

\section{Contributions from the three-graviton cut}
\label{sec:threecut}

The important difference with the maximal supergravity computation in~\cite{Bjerrum-Bohr:2021vuf} is that the two-loop amplitude involves integrals with
non-trivial numerators and with more topologies.
Using a partial fraction decomposition of the tree-level amplitudes
with respects to the linear propagators $p_1\cdot l_i$ and $p_2\cdot
l_i$ with $i=1,2,3$, one can reorganise the three-particle cut into five distinct topologies that will contribute to the classical result
\begin{equation}
\mathcal M_{2}^{\rm 3-cut}(\sigma,q^2)=\mathcal M_2^{\dBox}+\mathcal
M_2^{\triangleboxleft }+\mathcal M_2^{\triangleboxright}+\mathcal
M_2^{\triangleleft\triangleleft}+\mathcal M_2^{\triangleright\triangleright}+\mathcal
M_2^{H}+\mathcal M_2^{\Box\circ},
\end{equation}
with the result 
\begin{multline}\label{e:ResultM3cut}
\mathcal M_2^{\rm 3-cut(-1)}(\sigma,q^2)=\frac{2(4\pi
  e^{-\gamma_E})^{2\epsilon} \pi G_N^3 m_1^2 m_2^2 }{3
   \epsilon|\underline q|^{4\epsilon}
  \hbar}\Bigg(\frac{3s(2\sigma^2-1)^3}{(\sigma^2-1)^2}\cr
+\frac{im_1 m_2(2\sigma^2-1)}{\pi \epsilon
  (\sigma^2-1)^{\frac{3}{2}}}\left(\frac{1-49\sigma^2+18
    \sigma^4}{5}-\frac{6 \sigma(2 \sigma^2-1)(6
    \sigma^2-7)\arccosh(\sigma)}{\sqrt{\sigma^2-1}}\right)\cr
-\frac{9(2\sigma^2-1)(1-5\sigma^2)s}{2(\sigma^2-1)}+\frac{3}{2}(m_1^2+m_2^2)(-1+18\sigma^2)-m_1
m_2\sigma(103+2\sigma^2)\cr
+\frac{12 m_1 m_2(3+12
  \sigma^2-4\sigma^4)\arccosh(\sigma)}{\sqrt{\sigma^2-1}} \cr
-\frac{6i m_1 m_2(2\sigma^2-1)^2}{\pi \epsilon \sqrt{\sigma^2-1}} \left(\frac{-1}{4(\sigma^2-1)}\right)^{\epsilon}\frac{d}{d\sigma} \left(\frac{(2\sigma^2-1)\arccosh(\sigma)}{\sqrt{\sigma^2-1}} \right)\Bigg).
\end{multline}

The expressions for the numerator factors of each of these integrals
are given in Appendix~\ref{sec:numerators}.
The  double-box integrals $\mathcal M_2^{\dBox}$ are evaluated in section~\ref{sec:double-box-contr},   the
box-triangles $\mathcal
M_2^{\triangleboxleft }$ and $\mathcal M_2^{\triangleboxright}$
are evaluated in section~\ref{sec:boxtriangle}, the double-triangle integrals $\mathcal M_2^{\triangleright\triangleright}$ and $\mathcal
M_2^{\triangleleft\triangleleft}$ are evaluated in section~\ref{sec:triangle},
the $H$-diagram contributions $\mathcal
M_2^{H}$ are evaluated in section~\ref{sec:Hdiagram}, and
the box-bubble contributions $\mathcal
M_2^{\Box\circ}$ are evaluated in section~\ref{sec:box}.

\subsection{Double-box contributions}
\label{sec:double-box-contr}
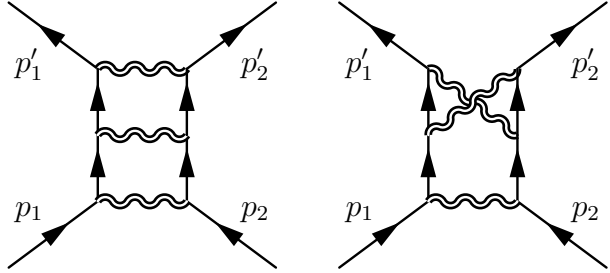
\begin{figure}[ht]
  \centering
  \begin{tabular}{cc}
    \begin{fmffile}{doublebox}
    \begin{fmfgraph*}(100,100)
\fmfstraight
\fmfleftn{i}{2}
\fmfrightn{o}{2}
\fmf{fermion,label=$p_1$,label.side=left}{i1,v1}
\fmf{fermion,label=$p_1'$,label.side=left}{v2,i2}
\fmf{fermion,label=$p_2'$,label.side=right}{v3,o2}
\fmf{fermion,label=$p_2$,label.side=right}{o1,v4}
\fmf{fermion}{v1,m1}
\fmf{fermion}{m1,v2}
\fmf{dbl_wiggly}{v2,v3}
\fmf{fermion}{m3,v3}
\fmf{fermion}{v4,m3}
\fmf{dbl_wiggly,tension=0}{m1,m3}
\fmf{phantom,tension=0}{v2,m3}
\fmf{phantom,tension=0}{m1,v3}
\fmf{dbl_wiggly}{v4,v1}
\end{fmfgraph*}
\end{fmffile}
&
    \begin{fmffile}{crossdoublebox}
    \begin{fmfgraph*}(100,100)
\fmfstraight
\fmfleftn{i}{2}
\fmfrightn{o}{2}
\fmf{fermion,label=$p_1$,label.side=left}{i1,v1}
\fmf{fermion,label=$p_1'$,label.side=left}{v2,i2}
\fmf{fermion,label=$p_2'$,label.side=right}{v3,o2}
\fmf{fermion,label=$p_2$,label.side=right}{o1,v4}
\fmf{fermion}{v1,m1}
\fmf{fermion}{m1,v2}
\fmf{phantom}{v2,v3}
\fmf{fermion}{m3,v3}
\fmf{fermion}{v4,m3}
\fmf{phantom,tension=0}{m1,m3}
\fmf{dbl_wiggly,tension=0}{v2,m3}
\fmf{dbl_wiggly,tension=0}{m1,v3}
\fmf{dbl_wiggly}{v4,v1}
\end{fmfgraph*}
\end{fmffile}
 \end{tabular}
  \caption{Double-box and crossed double-box diagrams.}
  \label{fig:doublebox}
\end{figure}
The double-box contributions arise from the sum of Feynman
graph topologies given in figure~\ref{fig:doublebox}. We provide the numerator
factors in Appendix~\ref{sec:numerators}.
Performing the tensorial
reductions with {\tt LiteRed}~\cite{Lee:2013mka}, we find that the double-box
contribution has the expansion

\begin{multline}
\mathcal M_2^{\dBox}(\sigma,q^2)=4096 \pi^3 G_N^3 m_1^5
m_2^5(2\sigma^2-1)^2\Big(m_1 m_2(2\sigma^2-1) (J_s+J_u)-6 \sigma
  \hbar^2 |\underline q|^2J_u\cr
  +8\sigma \hbar^2 |\underline q|^2 J_{\dBox}^{NP}\Big).
\end{multline}
The sum of  the integrals  $J_s+J_u$ has been evaluated in
section~4.3 of~\cite{Bjerrum-Bohr:2021vuf}, and the integral $J_u$ in
section~4.2 of~\cite{Bjerrum-Bohr:2021vuf}.

\paragraph{Evaluation of the  integral $J_{\dBox}^{NP}$}
This is a new contribution that did not appear in  the maximal
supergravity computation of ref.~\cite{Bjerrum-Bohr:2021vuf}. It reads (with
$D=4-2\epsilon$) 
\begin{multline} 
J_{\dBox }^{NP}=\frac{|\underline q|^{2D-10}}{16\hbar^3}\int \frac{d^D l_1
  d^D l_2}{(2\pi)^{2D}} \frac{l_2 \cdot l_3}{l_1^2 l_2^2
  (l_1+l_2+u_q)^2}\cr
\times\left(\frac{1}{(\bar{p}_1 \cdot
    l_1+i\varepsilon)(\bar{p}_1 \cdot l_2-i\varepsilon)}-\frac{1}{(\bar{p}_1
    \cdot l_2+i\varepsilon)(\bar{p}_1 \cdot
    l_1-i\varepsilon)}\right)\cr
\times
\left(\frac{1}{(\bar{p}_2 \cdot l_1-i\varepsilon)(\bar{p}_2 \cdot l_3+i\varepsilon)}-\frac{1}{(\bar{p}_2 \cdot l_3-i\varepsilon)(\bar{p}_2 \cdot l_1+i\varepsilon)}\right),
\end{multline}
where we have   scaled  the loop momenta
$l_i\to \hbar l_i\underline q$ as in~\cite{Bjerrum-Bohr:2021vuf}, and  defined
$p_1=\bar{p}_1+{\hbar\underline q\over2}$,
$p_2=\bar{p}_2-{\hbar\underline q\over2}$ and $q=|\vec q| u_q$ with $u_q^2=-1$.

Using the
 definition of the delta-function as a
distribution
\begin{equation}\label{e:PP}
 \lim_{\varepsilon\to0^+}\left( {1\over x-i\varepsilon}-{1\over x+i\varepsilon}\right)=
 \lim_{\varepsilon\to0^+} {2i\varepsilon \over x^2+\varepsilon^2}=2i\pi
\delta(x),
\end{equation}
the above expression can be written in terms of delta functions:
\begin{multline}
J_{\dBox }^{NP}=-\frac{|\underline q|^{2D-10}}{16\hbar^3}\int \frac{d^D l_1 d^D l_2}{(2\pi)^{2D-2}} \frac{l_2 \cdot l_3}{l_1^2 l_2^2 (l_1+l_2+u_q)^2}\left(\frac{\delta(\bar{p}_1 \cdot l_1)}{\bar{p}_1 \cdot l_2+i\varepsilon}-\frac{\delta(\bar{p}_1 \cdot l_2)}{\bar{p}_1 \cdot l_1+i\varepsilon}\right)\cr\times\left(\frac{\delta(\bar{p}_2 \cdot l_3)}{\bar{p}_2 \cdot l_1+i\varepsilon}-\frac{\delta(\bar{p}_2 \cdot l_1)}{\bar{p}_2 \cdot l_3+i\varepsilon}\right).
\end{multline}

We can therefore express the sum of all the double-box terms in Einstein
gravity in terms of integrals with delta-functions exactly as in the corresponding computation for maximal
supergravity~\cite{Herrmann:2021tct,Bjerrum-Bohr:2021vuf}.

Using {\tt LiteRed}~\cite{Lee:2013mka} we expand this expression  on the master
integrals used in~\cite{Bjerrum-Bohr:2021vuf} (see
appendix~\ref{sec:masters} for a summary of the results)\footnote{ The tensorial reduction is done after having localised  the
  integrals with the delta-function insertions. The tensorial reduction is then
  performed on the $D-1$ dimensional integrals.}
\begin{multline}
J_{\dBox }^{NP}=\frac{|\underline q|^{2D-10}}{96\hbar^3
  \epsilon^4 m_1^2 m_2^2(\sigma^2-1)}\left(\mathcal I_4(\sigma)+2
  \mathcal I_9(\sigma)\right)-\frac{|\underline q|^{2D-10}}{96
  \hbar^3\epsilon^4 m_1^2 m_2^2(\sigma^2-1)}\left(\mathcal
  I_4(\sigma)- \mathcal I_9(\sigma)\right)\cr
-\frac{|\underline q|^{2D-10}}{32\hbar^3}\frac{4(\arccosh(\sigma)-i\pi)}{m_1 m_2 \sqrt{\sigma^2-1}}\frac{i(4\pi e^{-\gamma_E})^{2\epsilon}}{128 m_1 m_2 \epsilon^2 \pi^3 \sqrt{\sigma^2-1}} ~.
\end{multline}
Next, using the evaluation of the master integral $\mathcal I_4(\sigma)$ in
section~5.2 and $\mathcal I_9(\sigma)$ in section~5.7
of ref.~\cite{Bjerrum-Bohr:2021vuf} we have
\begin{equation}
J_{\dBox }^{NP}=\frac{|\underline q|^{2D-10}(4\pi
  e^{-\gamma_E})^{2\epsilon}}{1024 \hbar^3\epsilon^2 \pi^2 m_1^2
  m_2^2(\sigma^2-1)}\left(\frac{i}{\pi}\left(-1\over4\right)^\epsilon
  \int_1^{\sigma} {dt\over (t^2-1)^{\frac{1}{2}+\epsilon}}-\frac{i}{\pi}\arccosh(\sigma)\right).
\end{equation}

\paragraph{The double-box}
Summing everything, the double-box contribution is given by 
\begin{equation}
  \mathcal M_2^{\dBox}(\sigma,q^2)={1\over |\underline
    q|^{4\epsilon}}\left(\mathcal M_2^{\dBox
    (-3)}(\sigma,q^2)+\mathcal M_2^{\dBox
    (-2)}(\sigma,q^2)+\mathcal M_2^{\dBox (-1)}(\sigma,q^2)+\mathcal O(\hbar)\right),
\end{equation}
where the superscript indicates the order of $\hbar$, with the leading term
\begin{equation}
\mathcal M_2^{\dBox(-3)}=-\frac{128\pi^3G_N^3m_1^4 m_2^4 (2\sigma^2-1)^3 \Gamma(-\epsilon)^3 \Gamma(1+2\epsilon)}{3\hbar^3|\underline q|^{2}(\sigma^2-1)(4\pi)^{2-2\epsilon} \Gamma(-3\epsilon)}.
\end{equation}
The first sub-leading piece is
\begin{equation}
\mathcal M_{2}^{\dBox(-2)}=\frac{256\pi^3G_N^3 i m_1^3 m_2^3 (m_1+m_2)(2\sigma^2-1)^3}{\hbar^2|\underline q| (\sigma^2-1)^{\frac{3}{2}}}\frac{\Gamma(\frac{1}{2}-\epsilon)^2 \Gamma(\frac{1}{2}+2\epsilon)\Gamma(-\epsilon) \Gamma(\frac{1}{2}-2\epsilon)}{(4\pi)^{\frac{5}{2}-2\epsilon}\Gamma(\frac{1}{2}-3\epsilon)\Gamma(-2\epsilon)},
\end{equation}
and the classical term, when written at leading orders in $\epsilon$,\footnote{For certain terms we keep $\epsilon$ in the exponent for reasons that will become clear below.} is
\begin{multline}
\mathcal M_{2}^{\dBox(-1)}= \frac{4G_N^3 m_1^3
  m_2^3(2\sigma^2-1)^2}{\hbar\pi}\left(4\pi
  e^{-\gamma_E} \right)^{2\epsilon} \Bigg(\frac{\pi^2
  s(2\sigma^2-1)}{2\epsilon
  m_1m_2(\sigma^2-1)^2}\cr
+\frac{i\pi((7-6\sigma^2)\sigma
  \arccosh(\sigma)-(2\sigma^2-1)\sqrt{\sigma^2-1})}{\epsilon^2
  (\sigma^2-1)^2}\cr
-\frac{i\pi}{\epsilon^2 \sqrt{\sigma^2-1}} \left(\frac{-1}{4(\sigma^2-1)}\right)^{\epsilon}\frac{d}{d\sigma} \Big(\frac{(2\sigma^2-1)\arccosh(\sigma)}{\sqrt{\sigma^2-1}} \Big)\Bigg).
\end{multline}
The first and the second lines agree with first quantum correction to
the one-loop box as given in ref.~\cite{KoemansCollado:2019ggb}. These pieces are
needed for the exponentiation of the eikonal phase to two-loop order.
The last line is the radiation-reaction term. We have written it compactly in terms of a $\sigma$-derivative. Notice that in Einstein
gravity this term is 
\begin{equation}
(2\sigma^2-1)^2\frac{d}{d\sigma}
\Big(\frac{(2\sigma^2-1)\arccosh(\sigma)}{\sqrt{\sigma^2-1}}\Big),
\end{equation}
whereas it is 
\begin{equation}\label{e:RRmaximal}
(2\sigma^2)^2\frac{d}{d\sigma}
\Big(\frac{2\sigma^2\arccosh(\sigma)}{\sqrt{\sigma^2-1}}\Big),
\end{equation}
in maximal supergravity. The difference is due to the exchange of the
dilaton field in the latter theory.

\subsection{Box-Triangle contributions}\label{sec:boxtriangle}
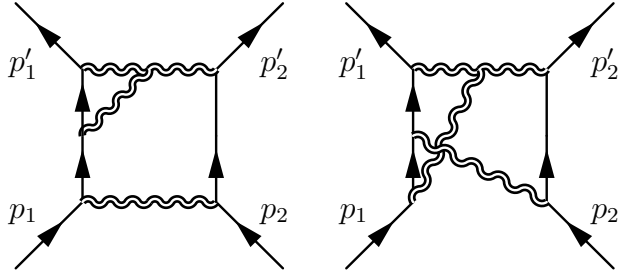
\begin{figure}[ht]
  \centering
  \begin{tabular}{cc}
    \begin{fmffile}{trianglebox}
    \begin{fmfgraph*}(100,100)
\fmfstraight
\fmfleftn{i}{2}
\fmfrightn{o}{2}
\fmf{fermion,label=$p_1$,label.side=left}{i1,v1}
\fmf{fermion,label=$p_1'$,label.side=left}{v2,i2}
\fmf{fermion,label=$p_2'$,label.side=right}{v3,o2}
\fmf{fermion,label=$p_2$,label.side=right}{o1,v4}
\fmf{dbl_wiggly,tension=1}{v2,m4}
\fmf{dbl_wiggly,tension=1}{m4,v3}
\fmf{dbl_wiggly,tension=1}{v4,m2}
\fmf{dbl_wiggly,tension=1}{m2,v1}
\fmf{fermion,tension=1}{v1,m1}
\fmf{fermion,tension=1}{m1,v2}
\fmf{plain,tension=1}{v3,m3}
\fmf{fermion,tension=1}{v4,m3}
\fmf{dbl_wiggly,tension=0}{m4,m1}
\fmf{phantom,tension=0}{m2,m3}
\fmf{phantom,tension=0}{m3,m4}
\fmf{phantom,tension=0}{m1,m2}
\end{fmfgraph*}
\end{fmffile}
&
    \begin{fmffile}{boxtriangle}
    \begin{fmfgraph*}(100,100)
\fmfstraight
\fmfleftn{i}{2}
\fmfrightn{o}{2}
\fmf{fermion,label=$p_1$,label.side=left}{i1,v1}
\fmf{fermion,label=$p_1'$,label.side=left}{v2,i2}
\fmf{fermion,label=$p_2'$,label.side=right}{v3,o2}
\fmf{fermion,label=$p_2$,label.side=right}{o1,v4}
\fmf{dbl_wiggly,tension=1}{v2,m4}
\fmf{dbl_wiggly,tension=1}{m4,v3}
\fmf{phantom,tension=1}{v4,m2}
\fmf{phantom,tension=1}{m2,v1}
\fmf{fermion,tension=1}{v1,m1}
\fmf{fermion,tension=1}{m1,v2}
\fmf{plain,tension=1}{v3,m3}
\fmf{fermion,tension=1}{v4,m3}
\fmf{phantom,tension=0}{m4,m1}
\fmf{phantom,tension=0}{m2,m3}
\fmf{phantom,tension=0}{m3,m4}
\fmf{phantom,tension=0}{m1,m2}
\fmf{dbl_wiggly,tension=0}{m4,v1}
\fmf{dbl_wiggly,tension=0}{m1,v4}
\end{fmfgraph*}
\end{fmffile}
 \end{tabular}
  \caption{Box-triangle graphs.}
  \label{fig:boxtriangle}
\end{figure}
The box-triangle contributions are given by the sum of Feynman
integrals topologies given in figure~\ref{fig:boxtriangle}, together with the mirrored
ones with the graviton line attached to the other scalar
line. The numerator
factors are again provided in Appendix~\ref{sec:numerators}. Using {\tt LiteRed}~\cite{Lee:2013mka} for
performing the tensorial reduction, and after evaluation of the various integrals, this contribution expands into
\begin{equation}
\mathcal M_2^{\triangleboxright}={1\over |\underline
  q|^{4\epsilon}}\left(
\mathcal M_{2}^{\triangleboxright(-2)}+\mathcal M_{2}^{\triangleboxright(-1)}
  +\mathcal O(\hbar^0)\right).
\end{equation}
There is no contribution of order $1/\hbar^3$. The leading order in the
$\hbar$ expansion is
\begin{equation}
\mathcal M_{2}^{\triangleboxright(-2)}=\frac{i6\pi^2 G_N^3m_1^4
  m_2^3(2\sigma^2-1)(1-5\sigma^2)(4\pi
  e^{-\gamma_E})^{2\epsilon}}{\epsilon \sqrt{\sigma^2-1}  \hbar^2
  |\underline q|}+\mathcal O(\epsilon^0),
\end{equation}
and the classical term is
\begin{multline}\label{e:M2other}
\mathcal M_{2}^{\triangleboxright(-1)}=(4\pi e^{-\gamma_E})^{2\epsilon}
{128\pi^3G_N^3m_1^3 m_2^2\over 3\hbar}\Bigg(
\frac{3im_2(2\sigma^2-1)(22\sigma^2-1)}{64 \epsilon^2 \pi^3
  \sqrt{\sigma^2-1}}\cr
-\frac{9(2\sigma^2-1)(1-5\sigma^2)(m_1+m_2 \sigma)}{128(\sigma^2-1)\epsilon \pi^2}-\frac{3(2\sigma^2-1)m_1}{128 \epsilon \pi^2}-\frac{m_2\sigma(55+2\sigma^2)}{128 \epsilon \pi^2} \Bigg)\,.
\end{multline}
The symmetric contribution $\mathcal M_2^{\triangleboxleft}$ is simply
obtained by the exchange of $m_1\leftrightarrow m_2$.

\subsection{Double-triangle contributions}\label{sec:triangle}

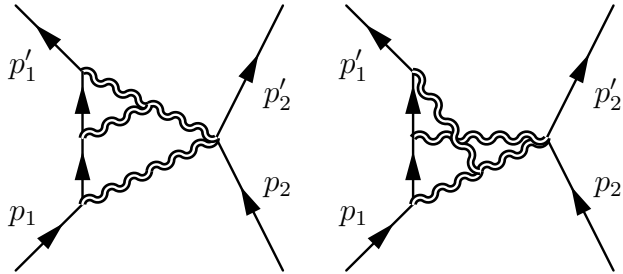
\begin{figure}[ht]
  \centering
  \begin{tabular}{cc}
    \begin{fmffile}{triangle}
    \begin{fmfgraph*}(100,100)
\fmfstraight
\fmfleftn{i}{2}
\fmfrightn{o}{2}
\fmf{fermion,label=$p_1$,label.side=left}{i1,v1}
\fmf{fermion,label=$p_1'$,label.side=left}{v2,i2}
\fmf{fermion,label=$p_2'$,label.side=right}{v3,o2}
\fmf{fermion,label=$p_2$,label.side=right}{o1,v3}
\fmf{fermion,tension=0.5}{v1,m1}
\fmf{fermion,tension=0.5}{m1,v2}
\fmf{dbl_wiggly,tension=1}{v2,m4}
\fmf{dbl_wiggly,tension=1}{m4,v3}
\fmf{dbl_wiggly,tension=1}{v3,m2}
\fmf{dbl_wiggly,tension=1}{m2,v1}
\fmf{dbl_wiggly,tension=0}{m4,m1}
\end{fmfgraph*}
\end{fmffile}
&
    \begin{fmffile}{crosstriangle}
    \begin{fmfgraph*}(100,100)
\fmfstraight
\fmfleftn{i}{2}
\fmfrightn{o}{2}
\fmf{fermion,label=$p_1$,label.side=left}{i1,v1}
\fmf{fermion,label=$p_1'$,label.side=left}{v2,i2}
\fmf{fermion,label=$p_2'$,label.side=right}{v3,o2}
\fmf{fermion,label=$p_2$,label.side=right}{o1,v3}
\fmf{fermion,tension=0.5}{v1,m1}
\fmf{fermion,tension=0.5}{m1,v2}
\fmf{phantom,tension=1}{v2,m4}
\fmf{phantom,tension=1}{m4,v3}
\fmf{dbl_wiggly,tension=1}{v3,m2}
\fmf{dbl_wiggly,tension=1}{m2,v1}
\fmf{phantom,tension=0}{m2,m1}
\fmf{dbl_wiggly,tension=0}{v3,m1}
\fmf{dbl_wiggly,tension=0}{m2,v2}
\end{fmfgraph*}
\end{fmffile}
 \end{tabular}
  \caption{Double-triangle graphs.}
  \label{fig:triangle}
\end{figure}

The double-triangle contributions are given by the sum of Feynman
integrals topologies given in figure~\ref{fig:triangle} together with the
mirrored ones
with the triangle attached to the other scalar line. We give the numerator
factors in Appendix~\ref{sec:numerators}.

We again use  {\tt LiteRed}~\cite{Lee:2013mka} for the tensorial reduction. Applying also the
identity in~\eqref{e:PP} we obtain the delta-function representation 
\begin{align}
\mathcal M_2^{\triangleright\triangleright}&=-{128 \pi^3G_N^3 m_1^6 m_2^2(10 \sigma^2-1) \over
  \hbar |\underline q|^{4\epsilon}}\int \frac{d^{D-1} l_1 d^{D-1}
                               l_2}{(2\pi)^{2D-2}}\frac{\delta(\bar{p}_1
                               \cdot l_1)\delta(\bar{p}_1\cdot
                               l_2)}{l_1^2 l_2^2 (l_1+l_2+u_q)^2},\cr
& =\frac{2\pi G_N^3 m_1^4 m_2^2(10 \sigma^2-1)(4\pi
  e^{-\gamma_E})^{2\epsilon}}{ \epsilon |\underline q|^{4\epsilon}\hbar },
\end{align}
and the symmetric triangle
\begin{equation}
\mathcal M_2^{\triangleleft\triangleleft}=\frac{2\pi G_N^3|\underline q|^{-4\epsilon}m_1^2
  m_2^4(10 \sigma^2-1)(4\pi e^{-\gamma_E})^{2\epsilon}}{
  \epsilon|\underline q|^{4\epsilon}\hbar }.
\end{equation}
Notice that the double-triangle integrals start contributing from the
classical order in the $\hbar\to0$ limit.

\subsection{The $H$-diagram}\label{sec:Hdiagram}
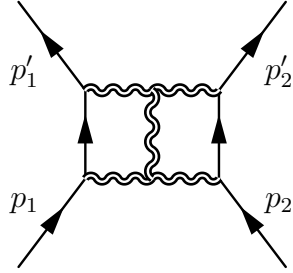
\begin{figure}[ht]
  \centering
    \begin{fmffile}{Hdiagram}
    \begin{fmfgraph*}(100,100)
\fmfstraight
\fmfleftn{i}{2}
\fmfrightn{o}{2}
\fmf{fermion,label=$p_1$,label.side=left}{i1,v1}
\fmf{fermion,label=$p_1'$,label.side=left}{v2,i2}
\fmf{fermion,label=$p_2'$,label.side=right}{v3,o2}
\fmf{fermion,label=$p_2$,label.side=right}{o1,v4}
\fmf{fermion}{v1,v2}
\fmf{fermion}{v4,v3}
\fmf{dbl_wiggly,tension=1}{v2,m4}
\fmf{dbl_wiggly,tension=1}{m4,v3}
\fmf{dbl_wiggly,tension=1}{v1,m2}
\fmf{dbl_wiggly,tension=1}{m2,v4}
\fmf{dbl_wiggly,tension=0}{m2,m4}
\end{fmfgraph*}
\end{fmffile}
  \caption{The $H$ diagram.}
  \label{fig:Hdiagram}
\end{figure}

The $H$-diagram integral is the $t$-channel integral similar to the
one evaluated in section~4.1 of~\cite{Bjerrum-Bohr:2021vuf} but this
time with tensorial numerator factor.
Reducing the numerator using {\tt LiteRed}~\cite{Lee:2013mka} we get
the expression
\begin{equation}
\mathcal M_2^{H}=-\frac{256\pi^3G_N^3m_1^3 m_2^3}{\sqrt{\sigma^2-1} |\underline q|^{4\epsilon}\hbar}\left(\frac{4(1+2\sigma^2)}{\epsilon^4} \mathcal I_2(\sigma)-\frac{4 \sigma \sqrt{\sigma^2-1}}{\epsilon^3}\mathcal I_4(\sigma)+\frac{(2\sigma^2-1)^2}{2\epsilon^4}\mathcal I_6(\sigma)\right)+\mathcal O(\epsilon^0).
\end{equation}
Using the evaluation of the master $\mathcal I_2(\sigma)$ and  $\mathcal
I_4(\sigma)$ in section~5.2, and $\mathcal I_6(\sigma)$ in section~5.6
in~\cite{Bjerrum-Bohr:2021vuf}  we obtain
\begin{equation}
\mathcal M_2^{H}=\frac{8\pi G_N^3 m_1^3 m_2^3(4\pi
  e^{-\gamma_E})^{2\epsilon}}{\epsilon \sqrt{\sigma^2-1}  |\underline
  q|^{4\epsilon}\hbar}((3+12\sigma^2-4\sigma^4) \arccosh(\sigma)-4
\sigma \sqrt{\sigma^2-1})+\mathcal O(\hbar^0).
\end{equation}
Notice that the $H$-diagram integral starts contributing from the
classical order in the $\hbar\to0$ limit. The full $H$-diagram has recently been considered in ref.~\cite{Kreer:2021sdt}.

\subsection{Box-bubble contribution}\label{sec:box}

\begin{figure}[h]
  \centering
    \begin{fmffile}{boxbubble}
   \begin{fmfgraph*}(100,100)
\fmfleftn{i}{2}
\fmfrightn{o}{2}
\fmf{fermion,label=$p_1$,label.side=left}{i1,v1}
\fmf{fermion,label=$p_1'$,label.side=left}{v2,i2}
\fmf{fermion,label=$p_2'$,label.side=right}{v3,o2}
\fmf{fermion,label=$p_2$,label.side=right}{o1,v4}
\fmf{fermion}{v1,v2}
\fmf{fermion}{v4,v3}
\fmf{dbl_wiggly,tension=1}{v2,m4}
\fmf{dbl_wiggly,tension=1}{m4,v3}
\fmf{dbl_wiggly,tension=1}{v1,m2}
\fmf{dbl_wiggly,left,tension=-0}{v2,v3}
\fmf{dbl_wiggly,tension=1}{m2,v4}
\fmf{phantom,tension=0}{m2,m4}
\end{fmfgraph*}
\end{fmffile}
  \caption{The box-bubble diagram.}
  \label{fig:box}
\end{figure}
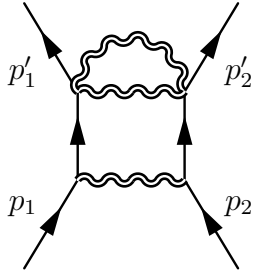

After tensorial reduction we find for the box-bubble numerator,  
\begin{equation}
\mathcal M_2^{\Box\circ}=-\frac{i 2G_N^3 m_1^3m_2^3(2\sigma^2-1)(1+522 \sigma^2)
 (4\pi e^{-\gamma_E})^{2\epsilon}}{15
  \epsilon^2 \sqrt{\sigma^2-1} |\underline q|^{4\epsilon}\hbar
}+\mathcal O(\hbar^0),
\end{equation}
which also contributes from the
classical order in the $\hbar\to0$ limit.

\section{Self-energy diagrams and vertex corrections}\label{sec:selfenergy}
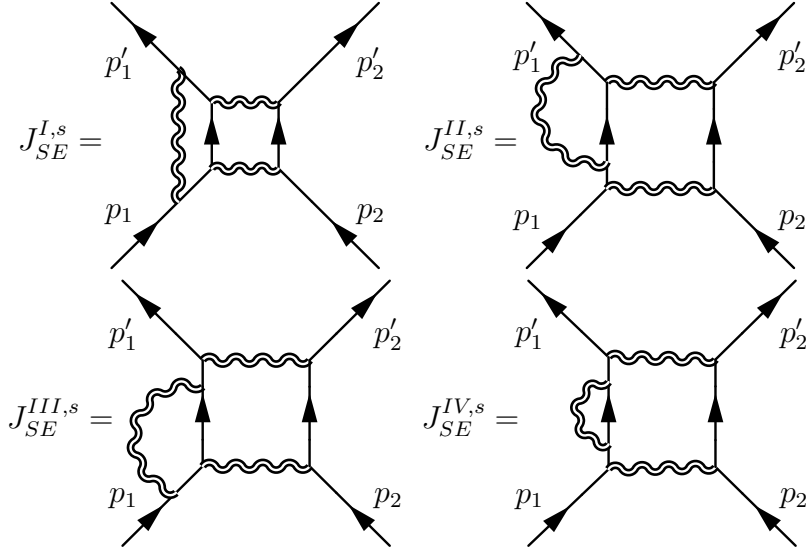
\begin{figure}[ht]
  \centering
  \begin{tabular}{cc}
$J_{SE}^{I,s}=\begin{gathered}
    \begin{fmffile}{SEI}
    \begin{fmfgraph*}(100,100)
\fmfstraight
\fmfleftn{i}{2}
\fmfrightn{o}{2}
\fmf{fermion,label=$p_1$,label.side=left}{i1,vv1}
\fmf{plain,tension=2}{vv1,v1}
\fmf{fermion,label=$p_1'$,label.side=left}{vv2,i2}
\fmf{plain,tension=2}{v2,vv2}
\fmf{fermion,label=$p_2'$}{vv3,o2}
\fmf{plain,tension=2}{v3,vv3}
\fmf{fermion,label=$p_2$,label.side=right}{o1,vv4}
\fmf{plain,tension=2}{v4,vv4}
\fmf{fermion}{v1,v2}
\fmf{fermion}{v4,v3}
\fmf{dbl_wiggly}{v4,v1}
\fmf{dbl_wiggly}{v2,v3}
\fmf{dbl_wiggly,tension=0}{vv1,vv2}
\fmf{phantom,tension=0}{vv3,vv4}
\end{fmfgraph*}
\end{fmffile}
\end{gathered}    $
    & 
      $J_{SE}^{II,s}= \begin{gathered} \begin{fmffile}{SEII}
    \begin{fmfgraph*}(100,100)
\fmfstraight
\fmfleftn{i}{2}
\fmfrightn{o}{2}
\fmf{fermion,label=$p_1$,label.side=left}{i1,vvv1}
\fmf{plain,tension=2}{vvv1,v1}
\fmf{fermion,label=$p_1'$,label.side=left}{vv2,i2}
\fmf{plain,tension=2}{v2,vv2}
\fmf{fermion,label=$p_2'$}{vv3,o2}
\fmf{plain,tension=2}{vv3,v3}
\fmf{fermion,label=$p_2$,label.side=right}{o1,vv4}
\fmf{plain,tension=2}{v4,vv4}
\fmf{dbl_wiggly,tension=0.5}{v4,v1}
\fmf{dbl_wiggly,tension=0.5}{v2,v3}
\fmf{plain,tension=2}{mm1,v2}
\fmf{fermion}{m1,mm1}
\fmf{plain,tension=2}{v1,m1}
\fmf{fermion}{mm3,m3}
\fmf{plain,tension=2}{v4,mm3}
\fmf{plain,tension=2}{v3,m3}
\fmf{dbl_wiggly,left,tension=0.}{m1,vv2}
\fmf{phantom,left,tension=0.}{vvv1,vv2}
\fmf{phantom,left,tension=0.}{m3,vv3}
\fmf{phantom,left,tension=0.}{m3,vv4}
\fmf{phantom,left,tension=0.}{m3,mm3}
\fmf{phantom,left,tension=0.}{m1,mm1}
\fmf{phantom,left,tension=0}{vvv1,vv2}
\fmf{phantom,left,tension=0}{vv3,vv4}
\end{fmfgraph*}
\end{fmffile}
    \end{gathered}$\\
  $J_{SE}^{III,s}=\begin{gathered} \begin{fmffile}{SEIII}
    \begin{fmfgraph*}(100,100)
\fmfstraight
\fmfleftn{i}{2}
\fmfrightn{o}{2}
\fmf{fermion,label=$p_1$,label.side=left}{i1,vvv1}
\fmf{plain,tension=2}{vvv1,v1}
\fmf{fermion,label=$p_1'$,label.side=left}{vv2,i2}
\fmf{plain,tension=2}{v2,vv2}
\fmf{fermion,label=$p_2'$}{vv3,o2}
\fmf{plain,tension=2}{vv3,v3}
\fmf{fermion,label=$p_2$,label.side=right}{o1,vv4}
\fmf{plain,tension=2}{v4,vv4}
\fmf{dbl_wiggly,tension=0.5}{v4,v1}
\fmf{dbl_wiggly,tension=0.5}{v2,v3}
\fmf{plain,tension=2}{mm1,v2}
\fmf{fermion}{m1,mm1}
\fmf{plain,tension=2}{v1,m1}
\fmf{fermion}{mm3,m3}
\fmf{plain,tension=2}{v4,mm3}
\fmf{plain,tension=2}{v3,m3}
\fmf{dbl_wiggly,left,tension=0.}{vvv1,mm1}
\fmf{phantom,left,tension=0.}{vvv1,vv2}
\fmf{phantom,left,tension=0.}{m3,vv3}
\fmf{phantom,left,tension=0.}{m3,vv4}
\fmf{phantom,left,tension=0.}{m3,mm3}
\fmf{phantom,left,tension=0.}{m1,mm1}
\fmf{phantom,left,tension=0}{vvv1,vv2}
\fmf{phantom,left,tension=0}{vv3,vv4}
\end{fmfgraph*}
\end{fmffile}
    \end{gathered}$&
  $J_{SE}^{IV,s}=\begin{gathered} \begin{fmffile}{SEIV}
    \begin{fmfgraph*}(100,100)
\fmfstraight
\fmfleftn{i}{2}
\fmfrightn{o}{2}
\fmf{fermion,label=$p_1$,label.side=left}{i1,vvv1}
\fmf{plain,tension=2}{vvv1,v1}
\fmf{fermion,label=$p_1'$,label.side=left}{vv2,i2}
\fmf{plain,tension=2}{v2,vv2}
\fmf{fermion,label=$p_2'$,label.side=right}{vv3,o2}
\fmf{plain,tension=2}{vv3,v3}
\fmf{fermion,label=$p_2$,label.side=right}{o1,vv4}
\fmf{plain,tension=2}{v4,vv4}
\fmf{dbl_wiggly,tension=0.5}{v4,v1}
\fmf{dbl_wiggly,tension=0.5}{v2,v3}
\fmf{plain,tension=2}{mm1,v2}
\fmf{fermion}{m1,mm1}
\fmf{plain,tension=2}{v1,m1}
\fmf{fermion}{mm3,m3}
\fmf{plain,tension=2}{v4,mm3}
\fmf{plain,tension=2}{v3,m3}
\fmf{phantom,left,tension=0.}{m1,vv2}
\fmf{phantom,left,tension=0.}{vvv1,vv2}
\fmf{phantom,left,tension=0.}{m3,vv3}
\fmf{phantom,left,tension=0.}{m3,vv4}
\fmf{phantom,left,tension=0.}{m3,mm3}
\fmf{dbl_wiggly,left,tension=-0.2}{m1,mm1}
\fmf{phantom,left,tension=0}{vvv1,vv2}
\fmf{phantom,left,tension=0}{vv3,vv4}
\end{fmfgraph*}
\end{fmffile}
    \end{gathered}$
  \end{tabular}
  \caption{Self-energy diagrams $s$-channel with the graviton line
    attached the one scalar line.}
  \label{fig:SelfEnergy}
\end{figure}

So far, all parts of the amplitude have followed from the three-graviton cut alone. This includes pieces that belong to both the conservative part
and the radiation-reaction part.
To get the full set of radiation-reaction contributions, we finally have  to compute the
two-loop integrals of self-energy and vertex corrections for the
massive lines that have not been included in the three-graviton cut. 
These contributions are not present in the maximal supergravity
amplitude because they are subleading in $\hbar\underline q$. In that
case, the radiation-reaction contribution is solely given by the
contribution in~\eqref{e:RRmaximal}.
They each have seven massive propagators (diagrams with  fever propagators
do not contribute to the  radiation-reaction, as they  are 
subleading in $\epsilon$). There are four independent seven-propagator diagrams listed in figure~\ref{fig:SelfEnergy}
in the $s$-channel with a graviton line attached to the scalar line
with mass $m_1$. There are four equivalent diagrams with the
graviton propagator attached to the other scalar line with mass $m_2$. There are as well
eight corresponding diagrams in what we can call the $u$-channel (corresponding to the crossed box).

The four associated integrals are 

\begin{multline}
J_{SE}^{I,s}= 4096\pi^3G_N^3 \hbar^7 \int \frac{d^D l_1 d^D l_2}{(2\pi)^{2D}}
\frac{
m_1^8 m_2^4 (2\sigma^2-1)^2+2m_1^6 m_2^4 (2\sigma^2-1)^2 |\hbar\underline q|^2
}{((p_1-l_1-l_2)^2-m_1^2+i\varepsilon)((p_1-l_2)^2-m_1^2+i\varepsilon)}\\ \times \frac{1}{((p_1-l_1-l_2-q)^2-m_1^2+i\varepsilon)((p_2-l_1)^2-m_2^2+i\varepsilon)l_1^2 (l_1+q)^2 (l_1+l_2)^2},
\end{multline}

\begin{multline}
J_{SE}^{II,s}= 4096\pi^3G_N^3 \hbar^7 \int \frac{d^D l_1 d^D l_2}{(2\pi)^{2D}}
\frac{
m_1^8 m_2^4 (2\sigma^2-1)^2
}{((p_1-l_1-l_2)^2-m_1^2+i\varepsilon)((p_1-l_2)^2-m_1^2+i\varepsilon)}\\ \times \frac{1}{((p_1+l_1)^2-m_1^2+i\varepsilon)((p_2-l_1)^2-m_2^2+i\varepsilon)l_1^2 (l_1+q)^2 (l_1+l_2)^2},
\end{multline}

\begin{multline}
  J_{SE}^{III,s}=  4096\pi^3G_N^3 \hbar^7\int \frac{d^D l_1 d^D l_2}{(2\pi)^{2D}} \frac{
m_1^8 m_2^4
  (2\sigma^2-1)^2
  }{((p_1+l_1)^2-m_1^2+i\varepsilon)((p_1-l_2)^2-m_1^2+i\varepsilon)}\\ \times \frac{1}{((p_1-l_1-l_2-q)^2-m_1^2+i\varepsilon)((p_2-l_1)^2-m_2^2+i\varepsilon)l_1^2 (l_1+q)^2 (l_1+l_2)^2},
\end{multline}

\begin{multline}
  J_{SE}^{IV,s}=  4096\pi^3G_N^3 \hbar^7\int \frac{d^D l_1 d^D l_2}{(2\pi)^{2D}} \frac{
m_1^8 m_2^4 (2\sigma^2-1)^2
  }{((p_1+l_1)^2-m_1^2+i\varepsilon)^2((p_1-l_2)^2-m_1^2+i\varepsilon)}\\ \times \frac{1}{((p_2-l_1)^2-m_2^2+i\varepsilon)l_1^2 (l_1+q)^2 (l_1+l_2)^2},
\end{multline}
with the numerators keeping only contributions that have weight
$m_2^4$ in $p_2$ because the other contributions are leading to
vanishing contributions having at least two delta-functions on $p_1\cdot l$
propagators. As well, only the part of the numerator that is
independent of the loop momenta contributes to the classical piece in the
$\epsilon$-expansion.
Summing all the contributions
\begin{equation}
\mathcal M_{SE}=-2^{16}\pi^3G_N^3\sum_{i=I}^{IV}
(J_{SE}^{i,s}+J_{SE}^{i,u})+(m_1\leftrightarrow m_2),
\end{equation}
the self-energy contribution is proportional  the master integrals
$\mathcal I_5(\sigma)$ evaluated in section~5
of~\cite{Bjerrum-Bohr:2021vuf} (see
appendix~\ref{sec:masters} for a summary of the results)
\begin{align}
\mathcal M_{SE}&=-\frac{1408\pi^3G_N^3m_1^3 m_2^3
  (2\sigma^2-1)^2}{3\epsilon^3\sqrt{\sigma^2-1} |\underline
                 q|^{4\epsilon}\hbar }\mathcal
                 I_5(\sigma)+\mathcal O(\epsilon^{-1})\cr
&=i\frac{44\pi G_N^3m_1^3 m_2^3
  (2\sigma^2-1)^2}{3\epsilon^2(\sigma^2-1)^{\epsilon+\frac12}
  |\underline q|^{4\epsilon}\hbar }\left(-1\over4\right)^\epsilon
                                                            (4\pi
                                                            e^{-\gamma_E})^{2\epsilon}+\mathcal
                                                            O(\epsilon^{-1})                 .
\end{align}

\section{The two-loop amplitude and the eikonal phase}

Summing up the three-particle cut and the self-energy diagrams and vertex corrections,
we obtain for  the total amplitude   

\begin{equation}\label{e:M2hbar}
  \mathcal M_2(\sigma, |\underline q|)={1\over |\underline q|^{4\epsilon}}\left( \mathcal M_2^{(-3)}(\sigma, |\underline q|)+
  \mathcal M_2^{(-2)}(\sigma, |\underline q|)+\mathcal
  M_2^{(-1)}(\sigma, |\underline q|)+\mathcal O(\hbar^0)\right),
\end{equation}
with the super-classical pieces
\begin{equation}
\mathcal M_{2}^{(-3)}(\sigma, |\underline q|)=-\frac{8 \pi G_N^3 m_1^4 m_2^4 (2\sigma^2-1)^3 \Gamma(-\epsilon)^3 \Gamma(1+2\epsilon)}{3\hbar^3|\underline q|^{2}(\sigma^2-1)(4\pi)^{-2\epsilon} \Gamma(-3\epsilon)},
\end{equation}
and
\begin{equation}
\mathcal M_{2}^{(-2)}(\sigma, |\underline q|)=\frac{6i \pi^2 G_N^3 (m_1+m_2)
  m_1^3 m_2^3(2\sigma^2-1)(1-5\sigma^2)(4\pi e^{-\gamma_E})^{2\epsilon}}{\epsilon
  \sqrt{\sigma^2-1}\hbar^2 |\underline q|}+\mathcal O(\epsilon^0),
\end{equation}
together with the classical term to the order $1/\epsilon$ for the real part
and $1/\epsilon^2$ for the imaginary terms
\begin{multline}
\mathcal M_{2}^{(-1)}(\sigma, |\underline q|)=\frac{2 \pi G_N^3(4\pi
  e^{-\gamma_E})^{2\epsilon} m_1^2 m_2^2 }{ \hbar
  \epsilon}\Bigg(\frac{s(2\sigma^2-1)^3}{(\sigma^2-1)^2}\cr
+\frac{im_1
  m_2(2\sigma^2-1)}{\pi \epsilon
  (\sigma^2-1)^{\frac{3}{2}}}\Big(\frac{1-49\sigma^2+18
  \sigma^4}{15}-\frac{2\sigma(7-20 \sigma^2+12
  \sigma^4)\arccosh(\sigma)}{\sqrt{\sigma^2-1}}\Big)\cr
-\frac{3(2\sigma^2-1)(1-5\sigma^2)s}{2(\sigma^2-1)}+\frac{1}{2}(m_1^2+m_2^2)(18\sigma^2-1)-\frac{1}{3}m_1
m_2\sigma(103+2\sigma^2)\cr
+\frac{4 m_1 m_2(3+12
  \sigma^2-4\sigma^4)\arccosh(\sigma)}{\sqrt{\sigma^2-1}} \cr
-\frac{2i m_1 m_2(2\sigma^2-1)^2}{\pi \epsilon \sqrt{\sigma^2-1}} \left(\frac{-1}{4(\sigma^2-1)}\right)^{\epsilon}\bigg(-\frac{11}{3}+\frac{d}{d\sigma} \Big(\frac{(2\sigma^2-1)\arccosh(\sigma)}{\sqrt{\sigma^2-1}} \Big) \bigg)\Bigg).
\end{multline}
The last line gives the radiation-reaction contributions
\begin{multline}\label{e:RR}
  \mathcal M_2^{(-1)}(\sigma, |\underline q|)\Big|_{\rm Rad.}=-\frac{4i G_N^3(4\pi
  e^{-\gamma_E})^{2\epsilon} m_1^3 m_2^3 }{\hbar
  \epsilon^2}\frac{ (2\sigma^2-1)^2}{\sqrt{\sigma^2-1}}
\left(\frac{-1}{4(\sigma^2-1)}\right)^{\epsilon}\cr
\times\bigg(-\frac{11}{3}+\frac{d}{d\sigma} \Big(\frac{(2\sigma^2-1)\arccosh(\sigma)}{\sqrt{\sigma^2-1}} \Big) \bigg).
\end{multline}
The $-11/3$
comes solely from the self-energy diagrams of
section~\ref{sec:selfenergy} and the derivative term from the
double-box diagrams of section~\ref{sec:double-box-contr}. 
The real and imaginary parts of this term clearly satisfies the relation
conjectured in~\cite{DiVecchia:2021ndb,DiVecchia:2021bdo}
\begin{equation}
\lim_{\epsilon\to0}  \epsilon\Re\eqref{e:RR}=- \lim_{\epsilon\to0}\epsilon^2\pi \Im\eqref{e:RR}.
\end{equation}

\subsection{The amplitude in $b$-space and eikonal exponentiation}

The amplitude is $b$-space is defined by
\begin{equation}
  \widetilde{\mathcal M}_2(\sigma,b)=\frac{1}{4 E_{\rm
      c.m.}P}\int_{\mathbb R^{D-2}} \frac{d^{D-2}\vec{\underline
      q}}{(2\pi)^{D-2}}\mathcal M_2(p_1,p_2,p_1',p_2') e^{i
    \vec{\underline q}\cdot\vec{b}}\,,
\end{equation}
where $4 E_{\rm c.m.}P= 4m_1 m_2 \sqrt{\sigma^2-1}$ and $E_{\rm c.m.}=\sqrt{s}$.

The two-loop amplitude in~\eqref{e:M2hbar} naturally decomposes as
follows after Fourier transform to $b$-space 
\begin{multline}\label{e:M2result}
\widetilde{  \mathcal M}_2(\sigma,b)=-\frac16
\left(\widetilde{\mathcal M}_0^{(-1)}(\sigma,b)\right)^3+ i
\widetilde{\mathcal M}_0^{(-1)}(\sigma,b)\left( \widetilde{\mathcal
  M}_1^{\rm Cl.}(\sigma,b)+ \widetilde{\mathcal
  M}_1^{\rm Qt.}(\sigma,b)\right)\cr+
\widetilde{\mathcal M}_{2}^{\rm  Cl.}(\sigma,b)+\mathcal O(\hbar^0).
\end{multline}
We note the following identifications, observed already at the level of diagram topologies:
\begin{align}\label{e:M2decomposition}
\widetilde{\mathcal M}_2^{\dBox(-3)}(\sigma,b)&= -{1\over6} \left(\widetilde{\mathcal
  M}_0^{(-1)}(\sigma,b) \right)^3,\cr
\widetilde{\mathcal M}_2^{\dBox(-2)}(\sigma,b)&= i\widetilde{\mathcal
  M}_0^{(-1)}(\sigma,b) \widetilde{\mathcal
                                                M}_1^{\Box(-1)}(\sigma,b),\cr
 \widetilde{\mathcal M}_2^{\triangleboxleft(-2)}(\sigma,b)+ \widetilde{\mathcal M}_2^{\triangleboxright(-2)}(\sigma,b)&= i \widetilde{\mathcal
  M}_0^{(-1)}(\sigma,b)\left( \widetilde{\mathcal
                                                                                                                                                   M}_1^{\triangleleft(-1)}(\sigma,b)+\widetilde{\mathcal
                                               M}_1^{\triangleright(-1)}(\sigma,b)\right),\cr                                               
\widetilde{\mathcal M}_2^{\dBox(-1)}(\sigma,b)&= i\widetilde{\mathcal
  M}_0^{(-1)}(\sigma,b) \widetilde{\mathcal
                                               M}_1^{\Box(0)}(\sigma,b)+\widetilde{\mathcal
                                               M}_2^{\dBox~{\rm Cl.}}(\sigma,b),\cr 
\widetilde{\mathcal M}_2^{\triangleboxleft(-1)}(\sigma,b)+\widetilde{\mathcal M}_2^{\triangleboxright(-1)}(\sigma,b)&= i \widetilde{\mathcal
  M}_0^{(-1)}(\sigma,b) \left(\widetilde{\mathcal
                                               M}_1^{\triangleleft(0)}(\sigma,b)+\widetilde{\mathcal
                                                                                                                    M}_1^{\triangleright(0)}(\sigma,b)\right)\cr
                                                                                                                    &+\widetilde{\mathcal
                                                                                                                    M}_2^{\triangleboxleft~{\rm
                                                                                                                    Cl.}}(\sigma,b)+\widetilde{\mathcal
                                                                                                                    M}_2^{\triangleboxright~{\rm
                                                                                                                    Cl.}}(\sigma,b),\cr
      \widetilde{\mathcal M}_2^{\Box\circ(-1)}(\sigma,b)&= i \widetilde{\mathcal
  M}_0^{(-1)}(\sigma,b) \widetilde{\mathcal
                                               M}_1^{\circ(0)}(\sigma,b)+\widetilde{\mathcal
                                               M}_2^{\Box\circ~{\rm
                                                             Cl.}}(\sigma,b),
\end{align}
where
\begin{equation}\label{e:Mzero}
\widetilde{\mathcal M_0}^{(-1)}(\sigma,b)=\frac{G_N m_1 m_2 (2\sigma^2-1)
  \Gamma(-\epsilon)}{\sqrt{\sigma^2-1}\hbar} (\pi b^2)^{\epsilon},
\end{equation}
is 
the first Post-Minkowskian contribution.

The various pieces from the one-loop amplitude $\mathcal M_1$ are
detailed in appendix~\ref{sec:oneloop}.  In the above expressions,
$\widetilde{\mathcal M}_1^{\Box(0)}(\sigma,b) $ is the Fourier
transform of first quantum contribution from the one-loop boxes
in~\eqref{e:Boxquantum},
$\widetilde{\mathcal M}_1^{\triangleright(-1)}(\sigma,b)$ is the
Fourier transform of the classical piece from the one-loop triangle
in~\eqref{e:triangleclassical} and
$\widetilde{\mathcal M}_1^{\triangleright(0)}(\sigma,b)$ is the
Fourier transform of the first quantum correction from the one-loop
triangle in~\eqref{e:trianglequantum} (likewise for
$\widetilde{\mathcal M}_1^{\triangleleft(-1)}(\sigma,b)$ and
$\widetilde{\mathcal M}_1^{\triangleleft(0)}(\sigma,b)$), and finally
$ \widetilde{\mathcal M}_1^{\circ(0)}(\sigma,b)$ is the Fourier
transform of~\eqref{e:bubble}.

It is striking how the above factorizations arise within graph
topologies. This will be  explained in section~\ref{sec:velocity} below.

Collecting the classical and the leading quantum pieces of the
one-loop amplitude   as in~\eqref{e:M1expand}, we obtain after Fourier
transform to $b$-space the classical piece 
\begin{equation}\label{e:Mone}
\widetilde{\mathcal M}_{1}^{\rm Cl.}(\sigma,b)=\frac{3 \pi G_N^2 (m_1+m_2) m_1 m_2 (5\sigma^2-1)}{4 b \sqrt{\sigma^2-1}\hbar} (\pi b^2 e^{\gamma_E})^{2\epsilon}+\mathcal O(\epsilon),
\end{equation}
and the leading quantum correction 
\begin{multline}
 \widetilde{\mathcal M}_{1}^{\rm Qt.}(\sigma,b)=\frac{G_N^2 (\pi b^2
   e^{\gamma_E})^{2\epsilon} }{b^2}\Bigg(\frac{i \epsilon
   s(2\sigma^2-1)^2}{(\sigma^2-1)^2}\cr
 -\frac{m_1 m_2 }{\pi (\sigma^2-1)^{\frac{3}{2}}}\Big(\frac{1-49\sigma^2+18 \sigma^4}{15}-\frac{2\sigma(2\sigma^2-1)(6\sigma^2-7)\arccosh(\sigma)}{\sqrt{\sigma^2-1}}\Big) \Bigg).
\end{multline}
The first and the last term in this expression matches the one
derived in~\cite{KoemansCollado:2019ggb}, the second term arises from
the contributions of the triangle and bubble in the Einstein gravity
one-loop amplitude as detailed in appendix~\ref{sec:oneloop}.\footnote{
We note that  the static limit of the second line matches the
quantum correction to the one-loop amplitude evaluated in~\cite{BjerrumBohr:2002kt,Bjerrum-Bohr:2013bxa}
\begin{equation}
\lim_{\sigma\to1}\frac{1}{2(\sigma^2-1)}\Big(\frac{1-49\sigma^2+18 \sigma^4}{15}-\frac{2\sigma(2\sigma^2-1)(6\sigma^2-7)\arccosh(\sigma)}{\sqrt{\sigma^2-1}}\Big) = -\frac{41}{10}\,.
\end{equation}}

Finally,  we have defined  the classical third Post-Minkowskian
contribution\footnote{The classical part $\widetilde{\mathcal
                                                  M}_2^{\Box\circ~{\rm
                                                  Cl.}}(\sigma,b) $
                                              vanishes and 
                                              does contribute to this
                                              result. The amplitude $\widetilde{\mathcal
                                                  M}_2^{\Box\circ(-1)}(\sigma,b) $
                                              is only needed for the
                                              eikonalisation of the
                                              two-loop amplitude as
                                              shown in eq.~\eqref{e:Mboxcir}.}
\begin{align}\label{e:M2classical}
  \widetilde{\mathcal M}_{2}^{\rm Cl.}(\sigma,b)&\equiv  \widetilde{\mathcal
    M}_2^{\dBox~{\rm Cl.}}(\sigma,b)+\widetilde{\mathcal
                                                                                                                    M}_2^{\triangleboxleft~{\rm
                                                                                                                    Cl.}}(\sigma,b)+\widetilde{\mathcal
                                                                                                                    M}_2^{\triangleboxright~{\rm
                                                                                                                    Cl.}}(\sigma,b) \cr
                                                  &
  +\widetilde{\mathcal M}_2^{\triangleleft\triangleleft~{\rm Cl.}}
    +\widetilde{\mathcal M}_2^{\triangleright\triangleright~{\rm
                                                  Cl.}}+\widetilde{\mathcal M}_2^{H~{\rm Cl.}}+\widetilde{\mathcal
    M}_2^{SE~{\rm Cl.}}+\widetilde{\mathcal
                                                  M}_2^{\Box\circ~{\rm
                                                  Cl.}}(\sigma,b)\\
                                                  &=\frac{G_N^3 m_1
                                                    m_2 (\pi b^2
  e^{\gamma_E})^{3\epsilon} }{\hbar b^2
  \sqrt{\sigma^2-1}}\Bigg(\frac{3(2\sigma^2-1)(5\sigma^2-1)s}{2(\sigma^2-1)}+\frac{s-2m_1m_2\sigma}{2}(18\sigma^2-1)\cr
&-\frac{1}{3}m_1
m_2\sigma(103+2\sigma^2)
+\frac{4 m_1 m_2(3+12
                                                                                                                          \sigma^2-4\sigma^4)\arccosh(\sigma)}{\sqrt{\sigma^2-1}} \cr
\nonumber                                                                                                                          &-\frac{2i m_1 m_2(2\sigma^2-1)^2}{\pi \epsilon \sqrt{\sigma^2-1}} \left(\frac{-1}{4(\sigma^2-1)}\right)^{\epsilon}\bigg(-\frac{11}{3}+\frac{d}{d\sigma} \Big(\frac{(2\sigma^2-1)\arccosh(\sigma)}{\sqrt{\sigma^2-1}} \Big) \bigg)\Bigg),
\end{align}
the last line of this expression is the radiation-reaction part which matches the result of~\cite{DiVecchia:2021bdo}.

\subsection{Relation to the world-line formalism: velocity cuts}\label{sec:velocity}

As we have remarked earlier, the two-loop amplitude can be organized
with integrals involving delta functions as in our earlier case of maximal
supergravity~\cite{Bjerrum-Bohr:2021vuf}.
There is a simple correspondence between the order in the
$\hbar$ expansion and the number of delta-function insertions. Namely,
symbolically we have the following pattern of delta-function insertions
\begin{align}
  \mathcal M_2^{(-3)}&\sim {1\over \hbar^3{\underline q}^2}\int
  \delta(p_1 \cdot l_1)  \delta(p_1 \cdot l_2)  \delta(p_2 \cdot l_1)  \delta(p_2 \cdot l_2),\\
  \mathcal M_2^{(-2)}&\sim {1\over \hbar^2\underline q}\int
 \left( \delta(p_1 \cdot l_1)  \delta(p_1 \cdot l_2)  \delta(p_2 \cdot l_1)+\delta(p_1 \cdot l_1)  \delta(p_2 \cdot l_1)  \delta(p_2 \cdot l_2)\right),\cr
 \nonumber \mathcal M_2^{(-1)}&\sim {1\over \hbar}\int \left(\delta(p_1 \cdot l_1)  \delta(p_2 \cdot l_1)+
                                \delta(p_1 \cdot l_1)  \delta(p_1 \cdot l_2)\right)\\
  &+ {1\over \hbar}\int \left(\delta(p_1 \cdot l_1)  \delta(p_2 \cdot l_2)+\delta(p_2 \cdot l_1)  \delta(p_2 \cdot l_2)\right).
\end{align}
We remark that only the
inverse powers of $\hbar$ can be  localised using delta-functions. The quantum
contributions, starting from the order $\hbar^0$, are not reducible to
$D-1$ dimensional integrals.   

  Because the delta functions project loop momenta onto external momenta, and hence external four-velocities, we shall
call the action of such delta functions {\em velocity cuts}. They are not true $D$-dimensional cuts but they share many features with genuine cuts,
especially after Fourier transformation into $b$-space. Indeed, these velocity cuts do not decouple the integrals in
momentum space but certain parts of them lead to factorized forms in
$b$-space. This amplitude factorization is important for the exponentiation
of the eikonal phase. As will be explained next, we can readily follow the diagrammatics of this factorization in $b$-space by tracing out how the cuts
are distributed. We indicate a velocity cut by a red dashed bar across a massive line.

For instance, two velocity cuts ($i.e.$, two delta-function insertions) in the following diagram
$$
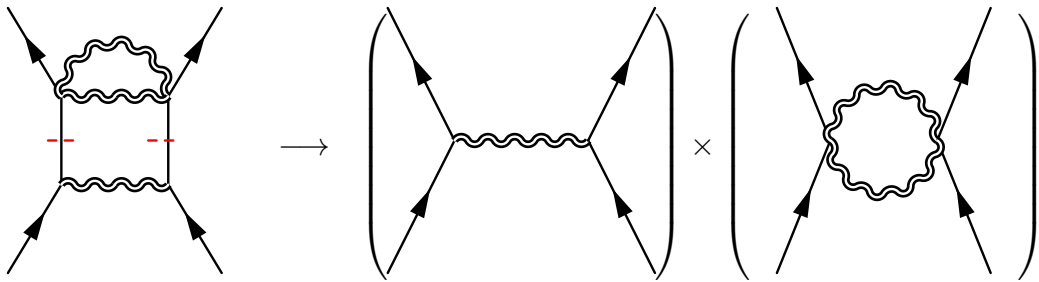
\begin{gathered}
    \begin{fmffile}{other10}
    \begin{fmfgraph*}(100,100)
\fmfleftn{i}{7}
\fmfrightn{o}{7}
\fmf{fermion}{i1,v1}
\fmf{fermion}{v2,i7}
\fmf{fermion}{v3,o7}
\fmf{fermion}{o1,v4}
\fmf{plain}{v1,v2}
\fmf{plain}{v4,v3}
\fmf{dbl_wiggly,tension=1}{v2,m4}
\fmf{dbl_wiggly,tension=1}{m4,v3}
\fmf{dbl_wiggly,tension=1}{v1,m2}
\fmf{phantom}{i4,h1}
\fmf{phantom}{h1,h2}
\fmf{phantom}{h2,h3}
\fmf{phantom}{h4,h3}
\fmf{phantom}{h4,y4}
\fmf{dashes,foreground=red}{y4,gg6}
\fmf{phantom}{gg6,ggg6}
\fmf{phantom}{ggg6,o4}
\fmf{phantom}{i4,hh1}
\fmf{phantom}{hh1,hhh1}
\fmf{dashes,foreground=red}{hhh1,hhh2}
\fmf{phantom}{hhh2,hh2}
\fmf{phantom}{hh2,hh3}
\fmf{phantom}{hh3,hh4}
\fmf{phantom}{hh4,hh5}
\fmf{phantom}{hh5,o4}
\fmf{dbl_wiggly,left,tension=-0}{v2,v3}
\fmf{dbl_wiggly,tension=1}{m2,v4}
\fmf{phantom,tension=0}{m2,m4}
\end{fmfgraph*}
\end{fmffile}
\end{gathered} \ \   \longrightarrow \ \ \left(\begin{gathered}
    \begin{fmffile}{other11}
    \begin{fmfgraph*}(100,100)
\fmfstraight
\fmfleftn{i}{2}
\fmfrightn{o}{2}
\fmf{fermion}{i1,v1}
\fmf{fermion}{v1,i2}
\fmf{fermion}{v3,o2}
\fmf{fermion}{o1,v3}
\fmf{fermion,tension=0.01}{v1,v2}
\fmf{fermion,tension=0.01}{v4,v3}
\fmf{dbl_wiggly}{v1,v3}
\end{fmfgraph*}
\end{fmffile}
\end{gathered}\right) \times \left(\begin{gathered}
    \begin{fmffile}{other12}
    \begin{fmfgraph*}(100,100)
\fmfleftn{i}{2}
\fmfrightn{o}{2}
\fmf{fermion}{i1,v1}
\fmf{fermion}{v1,i2}
\fmf{fermion}{v3,o2}
\fmf{fermion}{o1,v3}
\fmf{dbl_wiggly,left,tension=0.5}{v1,v3}
\fmf{dbl_wiggly,left,tension=0.5}{v3,v1}
\end{fmfgraph*}
\end{fmffile}
\end{gathered}\right)
$$
will in $b$-space lead to the product of the tree-level amplitude and
a one-loop bubble integral, so that the integral becomes 
\begin{align}\label{e:Mboxcir}
\widetilde{\mathcal M}_2^{\Box\circ(-1)}&=i \bigg(-\frac{G_N m_1
                                          m_2(2\sigma^2-1)(\pi
                                          b^2)^{\epsilon}}{\hbar\sqrt{\sigma^2-1}}
                                          \bigg)\times \bigg(\frac{G_N^2
                                          m_1 m_2 (1+522\sigma^2)(\pi
                                          b^2)^{2\epsilon}}{15\pi b^2
                                          \sqrt{\sigma^2-1}}\bigg)\cr
                                          &=i\widetilde{\mathcal
                                            M}_0^{(-1)}(\sigma,b)\times\widetilde{\mathcal M}_1^{\circ(0)}(\sigma,b),
\end{align}
since $\widetilde{\mathcal M}_2^{\Box\circ~{\rm Cl.}}(\sigma,b)$
vanishes because the two velocity-cuts are taken regarding the same
internal momenta.

Two velocity cuts of the box-triangle integral in figure~\ref{fig:deltaboxtriangle} lead, after a Fourier transform to $b$-space, to a
product of the tree-level amplitude and a triangle integral in exactly
the same way,\footnote{In
  particular we see the correspondence between the first line
  of~\eqref{e:M2other} and the quantum correction from the one-loop
  triangle in~\eqref{e:trianglequantum}.} plus 
an extra term contributing to the eikonal phase.
As one final example, let us consider two velocity cuts in the
double-box integral in figure~\ref{fig:deltadoublebox}. After a Fourier transformation to $b$-space, this leads to the product
of a tree-level amplitude times a one-loop box, in addition to the extra pieces shown.

\begin{figure}[ht]
    \begin{multline*}
    \begin{gathered}
    \begin{fmffile}{otherxx13}
    \begin{fmfgraph*}(100,100)
\fmfstraight
\fmfleftn{i}{2}
\fmfrightn{o}{2}
\fmf{plain}{i1,v1}
\fmf{plain}{v2,i2}
\fmf{plain}{v3,o2}
\fmf{plain}{o1,v4}
\fmf{dbl_wiggly,tension=1}{v2,m4}
\fmf{dbl_wiggly,tension=1}{m4,v3}
\fmf{dbl_wiggly,tension=1}{v4,m2}
\fmf{dbl_wiggly,tension=1}{m2,v1}
\fmf{plain,tension=1}{v1,m1}
\fmf{plain,tension=1}{m1,v2}
\fmf{plain,tension=1}{v3,m3}
\fmf{plain,tension=1}{v4,m3}
\fmf{dbl_wiggly,tension=0}{m4,m1}
\fmf{phantom,tension=0}{m2,m3}
\fmf{phantom,tension=0}{m3,m4}
\fmf{phantom,tension=0}{m1,m2}
\end{fmfgraph*}
\end{fmffile}
\end{gathered}    \longrightarrow 
\begin{gathered}
    \begin{fmffile}{otherxx14}
    \begin{fmfgraph*}(100,100)
\fmfstraight
\fmfleftn{i}{7}
\fmfrightn{o}{7}
\fmf{plain}{i1,v1}
\fmf{plain}{v2,i7}
\fmf{plain}{v3,o7}
\fmf{plain}{o1,v4}
\fmf{dbl_wiggly,tension=1}{v2,m4}
\fmf{dbl_wiggly,tension=1}{m4,v3}
\fmf{dbl_wiggly,tension=1}{v4,m2}
\fmf{dbl_wiggly,tension=1}{m2,v1}
\fmf{plain,tension=1}{v1,m1}
\fmf{plain,tension=1}{m1,v2}
\fmf{plain,tension=1}{m3,v3}
\fmf{plain,tension=1}{v4,m3}
\fmf{phantom}{i4,h1}
\fmf{phantom}{h1,h2}
\fmf{phantom}{h2,h3}
\fmf{phantom}{h3,h4}
\fmf{dashes,foreground=red}{h4,h5}
\fmf{phantom}{h5,o4}
\fmf{phantom}{i3,hh1}
\fmf{dashes,foreground=red}{hh1,hh2}
\fmf{phantom}{hh2,hh3}
\fmf{phantom}{hh3,hh4}
\fmf{phantom}{hh4,hh5}
\fmf{phantom}{hh5,o3}
\fmf{dbl_wiggly,tension=0}{m4,m1}
\fmf{phantom,tension=0}{m2,m3}
\fmf{phantom,tension=0}{m3,m4}
\fmf{phantom,tension=0}{m1,m2}
\end{fmfgraph*}
\end{fmffile}
\end{gathered} +  \begin{gathered}
    \begin{fmffile}{otherxx15}
    \begin{fmfgraph*}(100,100)
\fmfstraight
\fmfleftn{i}{7}
\fmfrightn{o}{7}
\fmf{plain}{i1,v1}
\fmf{plain}{v2,i7}
\fmf{plain}{v3,o7}
\fmf{plain}{o1,v4}
\fmf{dbl_wiggly,tension=1}{v2,m4}
\fmf{dbl_wiggly,tension=1}{m4,v3}
\fmf{dbl_wiggly,tension=1}{v4,m2}
\fmf{dbl_wiggly,tension=1}{m2,v1}
\fmf{plain,tension=1}{v1,m1}
\fmf{plain,tension=1}{m1,v2}
\fmf{plain,tension=1}{m3,v3}
\fmf{plain,tension=1}{v4,m3}
\fmf{phantom}{i4,h1}
\fmf{phantom}{h1,h2}
\fmf{phantom}{h2,h3}
\fmf{phantom}{h3,h4}
\fmf{dashes,foreground=red}{h4,h5}
\fmf{phantom}{h5,o4}
\fmf{phantom}{i5,hh1}
\fmf{dashes,foreground=red}{hh1,hh2}
\fmf{phantom}{hh2,hh3}
\fmf{phantom}{hh3,hh4}
\fmf{phantom}{hh4,hh5}
\fmf{phantom}{hh5,o5}
\fmf{dbl_wiggly,tension=0}{m4,m1}
\fmf{phantom,tension=0}{m2,m3}
\fmf{phantom,tension=0}{m3,m4}
\fmf{phantom,tension=0}{m1,m2}
\end{fmfgraph*}
\end{fmffile}
\end{gathered}+ \begin{gathered}
    \begin{fmffile}{otherxx150}
    \begin{fmfgraph*}(100,100)
\fmfstraight
\fmfleftn{i}{7}
\fmfrightn{o}{7}
\fmf{plain}{i1,v1}
\fmf{plain}{v2,i7}
\fmf{plain}{v3,o7}
\fmf{plain}{o1,v4}
\fmf{dbl_wiggly,tension=1}{v2,m4}
\fmf{dbl_wiggly,tension=1}{m4,v3}
\fmf{dbl_wiggly,tension=1}{v4,m2}
\fmf{dbl_wiggly,tension=1}{m2,v1}
\fmf{plain,tension=1}{v1,m1}
\fmf{plain,tension=1}{m1,v2}
\fmf{plain,tension=1}{m3,v3}
\fmf{plain,tension=1}{v4,m3}
\fmf{phantom}{i5,hh1}
\fmf{dashes,foreground=red}{hh1,hh2}
\fmf{phantom}{hh2,hh3}
\fmf{phantom}{hh3,hh4}
\fmf{phantom}{hh4,hh5}
\fmf{phantom}{hh5,o5}
\fmf{phantom}{i3,hhh1}
\fmf{dashes,foreground=red}{hhh1,hhh2}
\fmf{phantom}{hhh2,hhh3}
\fmf{phantom}{hhh3,hhh4}
\fmf{phantom}{hhh4,hhh5}
\fmf{phantom}{hhh5,o3}
\fmf{dbl_wiggly,tension=0}{m4,m1}
\fmf{phantom,tension=0}{m2,m3}
\fmf{phantom,tension=0}{m3,m4}
\fmf{phantom,tension=0}{m1,m2}
\end{fmfgraph*}
\end{fmffile}
\end{gathered} \cr
\longrightarrow
\left(\begin{gathered}
   \begin{fmffile}{otherxx16}
    \begin{fmfgraph*}(100,100)
\fmfstraight
\fmfleftn{i}{7}
\fmfrightn{o}{7}
\fmf{plain}{i1,v1}
\fmf{plain}{v2,i7}
\fmf{plain}{v3,o7}
\fmf{plain}{o1,v3}
\fmf{dbl_wiggly,tension=0.3}{v2,v4}
\fmf{dbl_wiggly,tension=0.3}{v1,v4}
\fmf{dbl_wiggly,tension=0.8}{v3,v4}
\fmf{plain}{v1,v2}
\end{fmfgraph*}
\end{fmffile}
\end{gathered}\ \right)\times\left(\begin{gathered}
    \begin{fmffile}{otherxx17}
    \begin{fmfgraph*}(100,100)
\fmfstraight
\fmfleftn{i}{7}
\fmfrightn{o}{7}
\fmf{plain}{i1,v1}
\fmf{plain}{v1,i7}
\fmf{plain}{v3,o7}
\fmf{plain}{o1,v3}
\fmf{dbl_wiggly}{v1,v3}
\end{fmfgraph*}
\end{fmffile}
\end{gathered}\right)\cr +  
\begin{gathered}
    \begin{fmffile}{otherxx18}
    \begin{fmfgraph*}(100,100)
\fmfstraight
\fmfleftn{i}{7}
\fmfrightn{o}{7}
\fmf{plain}{i1,v1}
\fmf{plain}{v2,i7}
\fmf{plain}{v3,o7}
\fmf{plain}{o1,v4}
\fmf{dbl_wiggly,tension=1}{v2,m4}
\fmf{dbl_wiggly,tension=1}{m4,v3}
\fmf{dbl_wiggly,tension=1}{v4,m2}
\fmf{dbl_wiggly,tension=1}{m2,v1}
\fmf{plain,tension=1}{v1,m1}
\fmf{plain,tension=1}{m1,v2}
\fmf{plain,tension=1}{m3,v3}
\fmf{plain,tension=1}{v4,m3}
\fmf{phantom}{i4,h1}
\fmf{phantom}{h1,h2}
\fmf{phantom}{h2,h3}
\fmf{phantom}{h3,h4}
\fmf{dashes,foreground=red}{h4,h5}
\fmf{phantom}{h5,o4}
\fmf{phantom}{i5,hh1}
\fmf{dashes,foreground=red}{hh1,hh2}
\fmf{phantom}{hh2,hh3}
\fmf{phantom}{hh3,hh4}
\fmf{phantom}{hh4,hh5}
\fmf{phantom}{hh5,o5}
\fmf{dbl_wiggly,tension=0}{m4,m1}
\fmf{phantom,tension=0}{m2,m3}
\fmf{phantom,tension=0}{m3,m4}
\fmf{phantom,tension=0}{m1,m2}
\end{fmfgraph*}
\end{fmffile}
\end{gathered} + 
\begin{gathered}
    \begin{fmffile}{otherxxx150}
    \begin{fmfgraph*}(100,100)
\fmfstraight
\fmfleftn{i}{7}
\fmfrightn{o}{7}
\fmf{plain}{i1,v1}
\fmf{plain}{v2,i7}
\fmf{plain}{v3,o7}
\fmf{plain}{o1,v4}
\fmf{dbl_wiggly,tension=1}{v2,m4}
\fmf{dbl_wiggly,tension=1}{m4,v3}
\fmf{dbl_wiggly,tension=1}{v4,m2}
\fmf{dbl_wiggly,tension=1}{m2,v1}
\fmf{plain,tension=1}{v1,m1}
\fmf{plain,tension=1}{m1,v2}
\fmf{plain,tension=1}{m3,v3}
\fmf{plain,tension=1}{v4,m3}
\fmf{phantom}{i5,hh1}
\fmf{dashes,foreground=red}{hh1,hh2}
\fmf{phantom}{hh2,hh3}
\fmf{phantom}{hh3,hh4}
\fmf{phantom}{hh4,hh5}
\fmf{phantom}{hh5,o5}
\fmf{phantom}{i3,hhh1}
\fmf{dashes,foreground=red}{hhh1,hhh2}
\fmf{phantom}{hhh2,hhh3}
\fmf{phantom}{hhh3,hhh4}
\fmf{phantom}{hhh4,hhh5}
\fmf{phantom}{hhh5,o3}
\fmf{dbl_wiggly,tension=0}{m4,m1}
\fmf{phantom,tension=0}{m2,m3}
\fmf{phantom,tension=0}{m3,m4}
\fmf{phantom,tension=0}{m1,m2}
\end{fmfgraph*}
\end{fmffile}
\end{gathered}
\end{multline*}
\caption{Velocity cuts of the box-triangle diagrams.}
\label{fig:deltaboxtriangle}
\end{figure}
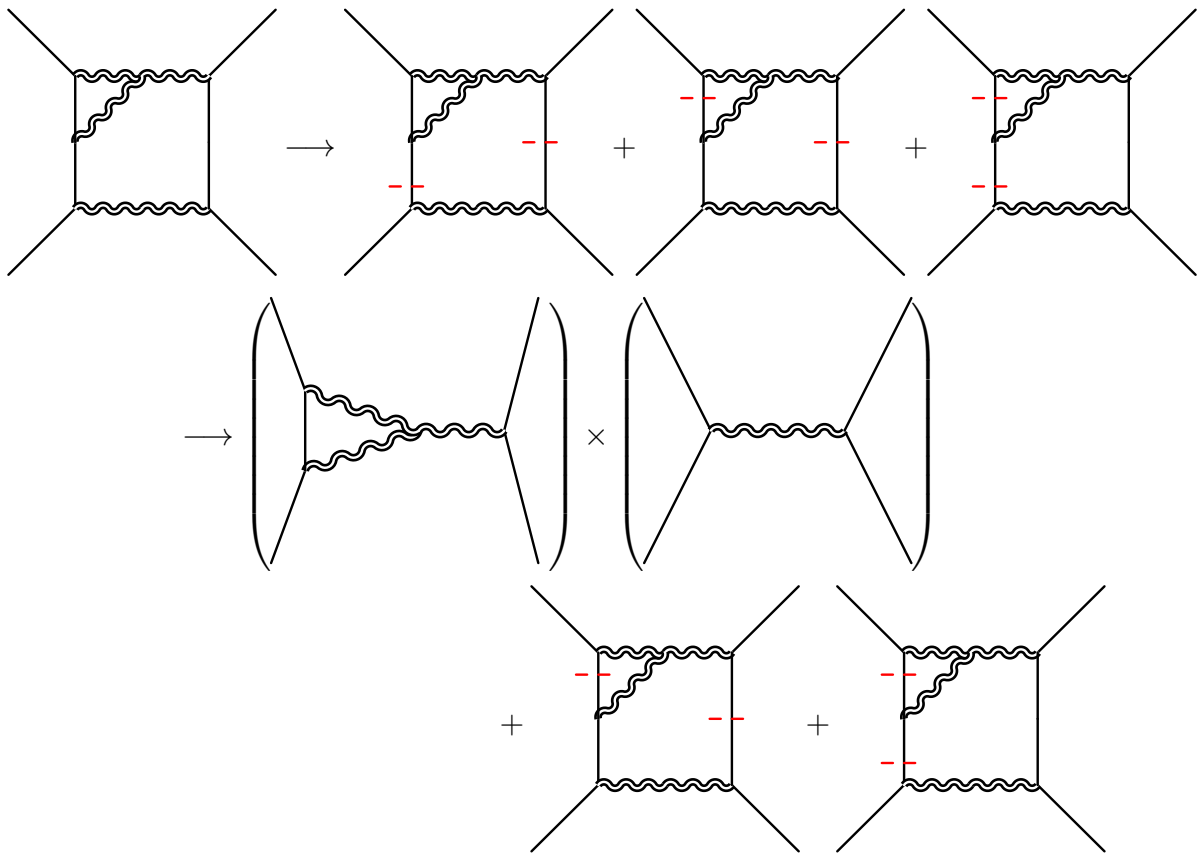

\begin{figure}[ht]
\begin{multline*}
    \begin{gathered}
    \begin{fmffile}{otherxx19x}
    \begin{fmfgraph*}(100,100)
\fmfstraight
\fmfleftn{i}{2}
\fmfrightn{o}{2}
\fmf{plain}{i1,v1}
\fmf{plain}{v2,i2}
\fmf{plain}{v3,o2}
\fmf{plain}{o1,v4}
\fmf{plain}{v1,m1}
\fmf{plain}{m1,v2}
\fmf{dbl_wiggly}{v2,v3}
\fmf{plain}{m3,v3}
\fmf{plain}{v4,m3}
\fmf{dbl_wiggly,tension=0}{m1,m3}
\fmf{phantom,tension=0}{v2,m3}
\fmf{phantom,tension=0}{m1,v3}
\fmf{dbl_wiggly}{v4,v1}
\end{fmfgraph*}
\end{fmffile}
\end{gathered} \longrightarrow 
\begin{gathered}
    \begin{fmffile}{otherxx20}
    \begin{fmfgraph*}(100,100)
\fmfstraight
\fmfleftn{i}{9}
\fmfrightn{o}{9}
\fmf{plain}{i1,v1}
\fmf{plain}{v2,i9}
\fmf{plain}{v3,o9}
\fmf{plain}{o1,v4}
\fmf{plain}{v1,m1}
\fmf{plain}{m1,v2}
\fmf{dbl_wiggly}{v2,v3}
\fmf{plain}{m3,v3}
\fmf{plain}{v4,m3}
\fmf{phantom}{i6,h1}
\fmf{phantom}{h1,h2}
\fmf{phantom}{h2,h3}
\fmf{phantom}{h3,g4}
\fmf{phantom}{g4,h4}
\fmf{dashes,foreground=red}{h4,h5}
\fmf{phantom}{h5,h6}
\fmf{phantom}{h6,o6}
\fmf{phantom}{i6,hh1}
\fmf{phantom}{hh1,hh2}
\fmf{dashes,foreground=red}{hh2,hh3}
\fmf{phantom}{hh3,hhhh4}
\fmf{phantom}{hhhh4,hh5}
\fmf{phantom}{hh5,hhh5}
\fmf{phantom}{hhh5,o6}
\fmf{dbl_wiggly,tension=0}{m1,m3}
\fmf{phantom,tension=0}{v2,m3}
\fmf{phantom,tension=0}{m1,v3}
\fmf{dbl_wiggly}{v4,v1}
\end{fmfgraph*}
\end{fmffile}
\end{gathered}
+\begin{gathered}
    \begin{fmffile}{otherxx20xx}
    \begin{fmfgraph*}(100,100)
\fmfstraight
\fmfleftn{i}{9}
\fmfrightn{o}{9}
\fmf{plain}{i1,v1}
\fmf{plain}{v2,i9}
\fmf{plain}{v3,o9}
\fmf{plain}{o1,v4}
\fmf{plain}{v1,m1}
\fmf{plain}{m1,v2}
\fmf{dbl_wiggly}{v2,v3}
\fmf{plain}{m3,v3}
\fmf{plain}{v4,m3}
\fmf{phantom}{i4,h1}
\fmf{phantom}{h1,h2}
\fmf{phantom}{h2,h3}
\fmf{phantom}{h3,g4}
\fmf{phantom}{g4,h4}
\fmf{dashes,foreground=red}{h4,h5}
\fmf{phantom}{h5,h6}
\fmf{phantom}{h6,o4}
\fmf{phantom}{i4,hh1}
\fmf{phantom}{hh1,hh2}
\fmf{dashes,foreground=red}{hh2,hh3}
\fmf{phantom}{hh3,hhhh4}
\fmf{phantom}{hhhh4,hh5}
\fmf{phantom}{hh5,hhh5}
\fmf{phantom}{hhh5,o4}
\fmf{dbl_wiggly,tension=0}{m1,m3}
\fmf{phantom,tension=0}{v2,m3}
\fmf{phantom,tension=0}{m1,v3}
\fmf{dbl_wiggly}{v4,v1}
\end{fmfgraph*}
\end{fmffile}
\end{gathered}\cr
+
\begin{gathered}
    \begin{fmffile}{otherxx21}
    \begin{fmfgraph*}(100,100)
\fmfstraight
\fmfleftn{i}{9}
\fmfrightn{o}{9}
\fmf{plain}{i1,v1}
\fmf{plain}{v2,i9}
\fmf{plain}{v3,o9}
\fmf{plain}{o1,v4}
\fmf{plain}{v1,m1}
\fmf{plain}{m1,v2}
\fmf{dbl_wiggly}{v2,v3}
\fmf{plain}{m3,v3}
\fmf{plain}{v4,m3}
\fmf{phantom}{i4,h1}
\fmf{phantom}{h1,h2}
\fmf{phantom}{h2,h3}
\fmf{phantom}{h3,g4}
\fmf{phantom}{g4,h4}
\fmf{dashes,foreground=red}{h4,h5}
\fmf{phantom}{h5,h6}
\fmf{phantom}{h6,o4}
\fmf{phantom}{i6,hh1}
\fmf{phantom}{hh1,hh2}
\fmf{dashes,foreground=red}{hh2,hh3}
\fmf{phantom}{hh3,hhhh4}
\fmf{phantom}{hhhh4,hh5}
\fmf{phantom}{hh5,hhh5}
\fmf{phantom}{hhh5,o6}
\fmf{dbl_wiggly,tension=0}{m1,m3}
\fmf{phantom,tension=0}{v2,m3}
\fmf{phantom,tension=0}{m1,v3}
\fmf{dbl_wiggly}{v4,v1}
\end{fmfgraph*}
\end{fmffile}
\end{gathered}+
\begin{gathered}
    \begin{fmffile}{otherxx21}
    \begin{fmfgraph*}(100,100)
\fmfstraight
\fmfleftn{i}{9}
\fmfrightn{o}{9}
\fmf{plain}{i1,v1}
\fmf{plain}{v2,i9}
\fmf{plain}{v3,o9}
\fmf{plain}{o1,v4}
\fmf{plain}{v1,m1}
\fmf{plain}{m1,v2}
\fmf{dbl_wiggly}{v2,v3}
\fmf{plain}{m3,v3}
\fmf{plain}{v4,m3}
\fmf{phantom}{i6,h1}
\fmf{phantom}{h1,h2}
\fmf{phantom}{h2,h3}
\fmf{phantom}{h3,g4}
\fmf{phantom}{g4,h4}
\fmf{dashes,foreground=red}{h4,h5}
\fmf{phantom}{h5,h6}
\fmf{phantom}{h6,o6}
\fmf{phantom}{i4,hh1}
\fmf{phantom}{hh1,hh2}
\fmf{dashes,foreground=red}{hh2,hh3}
\fmf{phantom}{hh3,hhhh4}
\fmf{phantom}{hhhh4,hh5}
\fmf{phantom}{hh5,hhh5}
\fmf{phantom}{hhh5,o4}
\fmf{dbl_wiggly,tension=0}{m1,m3}
\fmf{phantom,tension=0}{v2,m3}
\fmf{phantom,tension=0}{m1,v3}
\fmf{dbl_wiggly}{v4,v1}
\end{fmfgraph*}
\end{fmffile}
\end{gathered}\cr
\longrightarrow
2\left(
\begin{gathered}
    \begin{fmffile}{otherxx22}
    \begin{fmfgraph*}(100,100)
\fmfstraight
\fmfleftn{i}{9}
\fmfrightn{o}{9}
\fmf{plain}{i1,v1}
\fmf{plain}{v2,i9}
\fmf{plain}{v3,o9}
\fmf{plain}{o1,v4}
\fmf{dbl_wiggly}{v2,v3}
\fmf{dbl_wiggly}{v1,v4}
\fmf{plain}{v3,v4}
\fmf{plain}{v1,v2}
\end{fmfgraph*}
\end{fmffile}
\end{gathered}
\right)\times\left(
\begin{gathered}
    \begin{fmffile}{otherxx23}
    \begin{fmfgraph*}(100,100)
\fmfstraight
\fmfleftn{i}{9}
\fmfrightn{o}{9}
\fmf{plain}{i1,v1}
\fmf{plain}{v1,i9}
\fmf{plain}{v3,o9}
\fmf{plain}{o1,v3}
\fmf{dbl_wiggly}{v1,v3}
\end{fmfgraph*}
\end{fmffile}
\end{gathered}
\right)\cr
+
\begin{gathered}
    \begin{fmffile}{otherxx24xxx}
    \begin{fmfgraph*}(100,100)
\fmfstraight
\fmfleftn{i}{9}
\fmfrightn{o}{9}
\fmf{plain}{i1,v1}
\fmf{plain}{v2,i9}
\fmf{plain}{v3,o9}
\fmf{plain}{o1,v4}
\fmf{plain}{v1,m1}
\fmf{plain}{m1,v2}
\fmf{dbl_wiggly}{v2,v3}
\fmf{plain}{m3,v3}
\fmf{plain}{v4,m3}
\fmf{phantom}{i6,h1}
\fmf{phantom}{h1,h2}
\fmf{phantom}{h2,h3}
\fmf{phantom}{h3,g4}
\fmf{phantom}{g4,h4}
\fmf{dashes,foreground=red}{h4,h5}
\fmf{phantom}{h5,h6}
\fmf{phantom}{h6,o6}
\fmf{phantom}{i4,hh1}
\fmf{phantom}{hh1,hh2}
\fmf{dashes,foreground=red}{hh2,hh3}
\fmf{phantom}{hh3,hhhh4}
\fmf{phantom}{hhhh4,hh5}
\fmf{phantom}{hh5,hhh5}
\fmf{phantom}{hhh5,o4}
\fmf{dbl_wiggly,tension=0}{m1,m3}
\fmf{phantom,tension=0}{v2,m3}
\fmf{phantom,tension=0}{m1,v3}
\fmf{dbl_wiggly}{v4,v1}
\end{fmfgraph*}
\end{fmffile}
\end{gathered}
+
\begin{gathered}
    \begin{fmffile}{otherxx24}
    \begin{fmfgraph*}(100,100)
\fmfstraight
\fmfleftn{i}{9}
\fmfrightn{o}{9}
\fmf{plain}{i1,v1}
\fmf{plain}{v2,i9}
\fmf{plain}{v3,o9}
\fmf{plain}{o1,v4}
\fmf{plain}{v1,m1}
\fmf{plain}{m1,v2}
\fmf{dbl_wiggly}{v2,v3}
\fmf{plain}{m3,v3}
\fmf{plain}{v4,m3}
\fmf{phantom}{i4,h1}
\fmf{phantom}{h1,h2}
\fmf{phantom}{h2,h3}
\fmf{phantom}{h3,g4}
\fmf{phantom}{g4,h4}
\fmf{dashes,foreground=red}{h4,h5}
\fmf{phantom}{h5,h6}
\fmf{phantom}{h6,o4}
\fmf{phantom}{i6,hh1}
\fmf{phantom}{hh1,hh2}
\fmf{dashes,foreground=red}{hh2,hh3}
\fmf{phantom}{hh3,hhhh4}
\fmf{phantom}{hhhh4,hh5}
\fmf{phantom}{hh5,hhh5}
\fmf{phantom}{hhh5,o6}
\fmf{dbl_wiggly,tension=0}{m1,m3}
\fmf{phantom,tension=0}{v2,m3}
\fmf{phantom,tension=0}{m1,v3}
\fmf{dbl_wiggly}{v4,v1}
\end{fmfgraph*}
\end{fmffile}
\end{gathered}
\end{multline*}
\caption{Velocity cuts of the double-box diagram.}
\label{fig:deltadoublebox}
\end{figure}
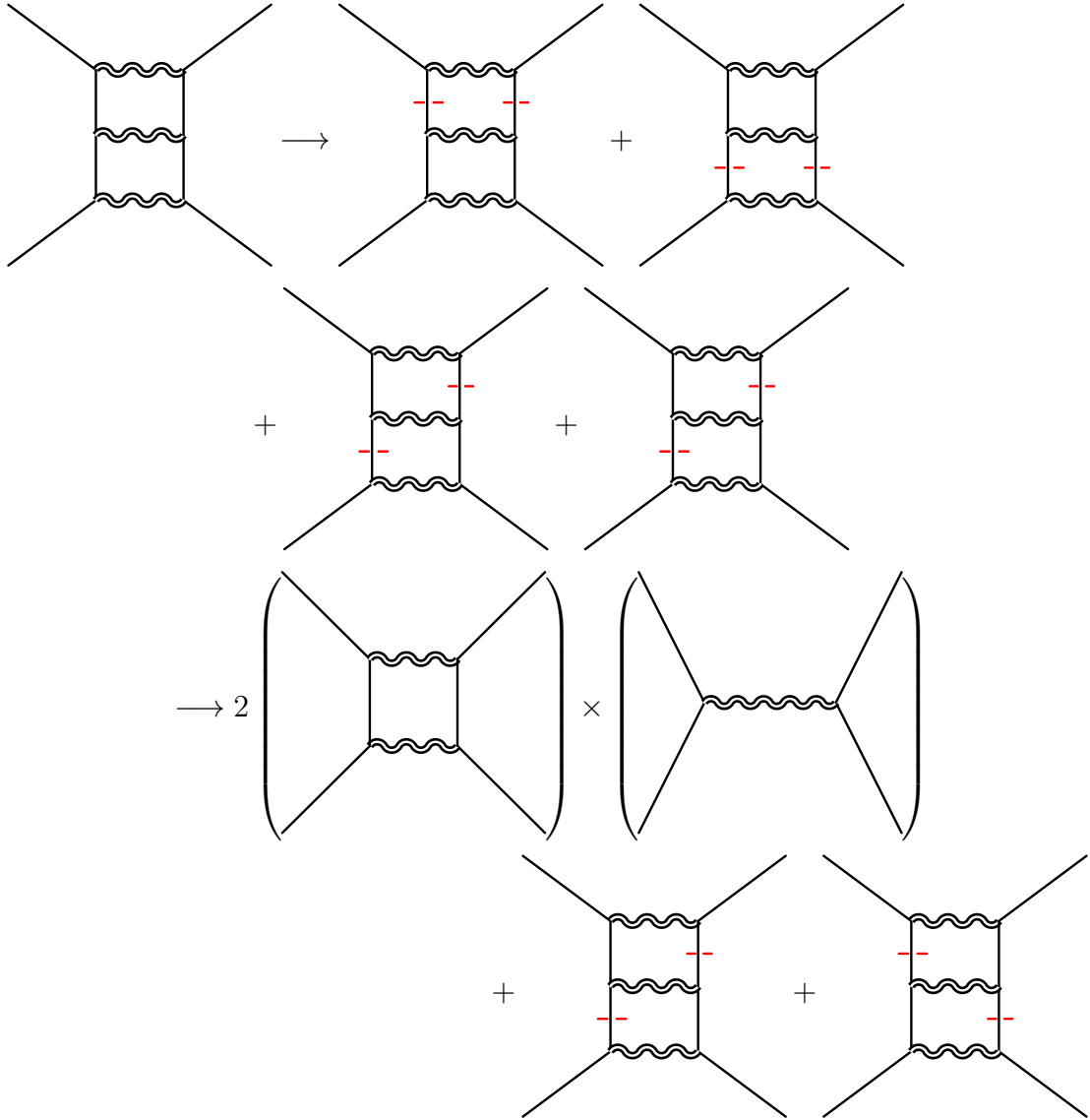

\begin{figure}[ht]
  \begin{tabular}{cc}
$\begin{gathered}
    \begin{fmffile}{newother1}
    \begin{fmfgraph*}(100,100)
\fmfleftn{i}{9}
\fmfrightn{o}{9}
\fmf{plain,label.side=left}{i1,v1}
\fmf{plain,label.side=left}{v2,i9}
\fmf{plain,label.side=right}{v3,o9}
\fmf{plain,label.side=right}{o1,v4}
\fmf{plain,tension=0.1}{v1,v2}
\fmf{plain,tension=0.1}{v3,v4}
\fmf{dbl_wiggly}{v2,m1}
\fmf{dbl_wiggly}{v4,m1}
\fmf{dbl_wiggly}{v3,m1}
\fmf{dbl_wiggly}{v1,m1}
\fmf{phantom}{i5,f2}
\fmf{phantom}{f2,f3}
\fmf{dashes,foreground=red}{f3,f4}
\fmf{phantom}{f4,f5}
\fmf{phantom}{f5,f6}
\fmf{dashes,foreground=red}{f6,f7}
\fmf{phantom}{f7,f8}
\fmf{phantom}{f8,o5}
\end{fmfgraph*}
\end{fmffile}
\end{gathered} \longrightarrow
\begin{gathered}
    \begin{fmffile}{newother2}
    \begin{fmfgraph*}(100,100)
\fmfleftn{i}{9}
\fmfrightn{o}{9}
\fmf{phantom}{i1,vv1}
\fmf{phantom}{vv2,i9}
\fmf{phantom}{vv3,o9}
\fmf{phantom}{o1,vv4}
\fmf{dbl_wiggly}{vv1,v1}
\fmf{dbl_wiggly}{v1,vv2}
\fmf{dbl_wiggly}{v1,vv3}
\fmf{dbl_wiggly}{vv4,v1}
\fmfv{decor.shape=circle,decor.filled=hatched, decor.size=3thick}{vv1}
\fmfv{decor.shape=circle,decor.filled=hatched, decor.size=3thick}{vv2}
\fmfv{decor.shape=circle,decor.filled=hatched, decor.size=3thick}{vv3}
\fmfv{decor.shape=circle,decor.filled=hatched, decor.size=3thick}{vv4}
\end{fmfgraph*}
\end{fmffile}
\end{gathered} $&
      $\begin{gathered}
    \begin{fmffile}{newother3}
    \begin{fmfgraph*}(100,100)
\fmfleftn{i}{9}
\fmfrightn{o}{9}
\fmf{plain,label.side=left}{i1,v1}
\fmf{plain,label.side=left}{v2,i9}
\fmf{plain,label.side=right}{v3,o9}
\fmf{plain,label.side=right}{o1,v4}
\fmf{plain,tension=1}{v1,v2}
\fmf{plain,tension=1}{v3,v4}
\fmf{dbl_wiggly}{v2,m1}
\fmf{dbl_wiggly}{v4,m3}
\fmf{dbl_wiggly}{v3,m1}
\fmf{dbl_wiggly}{v1,m3}
\fmf{dbl_wiggly,tension=0}{m1,m3}
\fmf{phantom}{i5,f2}
\fmf{phantom}{f2,f3}
\fmf{dashes,foreground=red}{f3,f4}
\fmf{phantom}{f4,f5}
\fmf{phantom}{f5,f6}
\fmf{dashes,foreground=red}{f6,f7}
\fmf{phantom}{f7,f8}
\fmf{phantom}{f8,o5}
\end{fmfgraph*}
\end{fmffile}
\end{gathered} \longrightarrow
\begin{gathered}
    \begin{fmffile}{newother4}
    \begin{fmfgraph*}(100,100)
\fmfleftn{i}{9}
\fmfrightn{o}{9}
\fmf{phantom}{i1,vv1}
\fmf{phantom}{vv2,i9}
\fmf{phantom}{vv3,o9}
\fmf{phantom}{o1,vv4}
\fmf{phantom}{vv1,vv2}
\fmf{phantom}{vv3,vv4}
\fmf{dbl_wiggly}{vv1,v3}
\fmf{dbl_wiggly}{v1,vv2}
\fmf{dbl_wiggly}{v1,vv3}
\fmf{dbl_wiggly}{vv4,v3}
\fmf{dbl_wiggly,tension=0}{v1,v3}
\fmfv{decor.shape=circle,decor.filled=hatched, decor.size=3thick}{vv1}
\fmfv{decor.shape=circle,decor.filled=hatched, decor.size=3thick}{vv2}
\fmfv{decor.shape=circle,decor.filled=hatched, decor.size=3thick}{vv3}
\fmfv{decor.shape=circle,decor.filled=hatched, decor.size=3thick}{vv4}
\end{fmfgraph*}
\end{fmffile}
\end{gathered} $\\
      $\begin{gathered}
    \begin{fmffile}{newother5}
    \begin{fmfgraph*}(100,100)
\fmfleftn{i}{9}
\fmfrightn{o}{9}
\fmf{plain,label.side=left}{i1,v1}
\fmf{plain,label.side=left}{v2,i9}
\fmf{plain,label.side=right}{v3,o9}
\fmf{plain,label.side=right}{o1,v4}
\fmf{plain,tension=0.3}{v1,v2}
\fmf{plain,tension=0.3}{v3,v4}
\fmf{dbl_wiggly}{v2,m1}
\fmf{dbl_wiggly}{v4,m3}
\fmf{dbl_wiggly}{v3,m3}
\fmf{dbl_wiggly}{v1,m1}
\fmf{dbl_wiggly,tension=2}{m1,m3}
\fmf{phantom}{i5,f2}
\fmf{phantom}{f2,f3}
\fmf{dashes,foreground=red}{f3,f4}
\fmf{phantom}{f4,ff5}
\fmf{phantom}{ff5,fff5}\fmf{phantom}{fff5,f5}
\fmf{phantom}{f5,f6}
\fmf{dashes,foreground=red}{f6,f7}
\fmf{phantom}{f7,f8}
\fmf{phantom}{f8,o5}
\end{fmfgraph*}
\end{fmffile}
\end{gathered} \longrightarrow
\begin{gathered}
    \begin{fmffile}{newother6}
    \begin{fmfgraph*}(100,100)
\fmfleftn{i}{9}
\fmfrightn{o}{9}
\fmf{phantom}{i1,vv1}
\fmf{phantom}{vv2,i9}
\fmf{phantom}{vv3,o9}
\fmf{phantom}{o1,vv4}
\fmf{phantom}{vv1,vv2}
\fmf{phantom}{vv3,vv4}
\fmf{dbl_wiggly}{vv1,v1}
\fmf{dbl_wiggly}{v1,vv2}
\fmf{dbl_wiggly}{v3,vv3}
\fmf{dbl_wiggly}{vv4,v3}
\fmf{dbl_wiggly}{v1,v3}
\fmfv{decor.shape=circle,decor.filled=hatched, decor.size=3thick}{vv1}
\fmfv{decor.shape=circle,decor.filled=hatched, decor.size=3thick}{vv2}
\fmfv{decor.shape=circle,decor.filled=hatched, decor.size=3thick}{vv3}
\fmfv{decor.shape=circle,decor.filled=hatched, decor.size=3thick}{vv4}
\end{fmfgraph*}
\end{fmffile}
\end{gathered} $&
      $\begin{gathered}
    \begin{fmffile}{newother7}
    \begin{fmfgraph*}(100,100)
\fmfleftn{i}{9}
\fmfrightn{o}{9}
\fmf{plain,label.side=left}{i1,v1}
\fmf{plain,label.side=left}{v2,i9}
\fmf{plain,label.side=right}{v3,o9}
\fmf{plain,label.side=right}{o1,v3}
\fmf{plain,tension=0.3}{v1,v5}
\fmf{plain,tension=0.3}{v2,v5}
\fmf{dbl_wiggly,tension=0}{v5,m1}
\fmf{dbl_wiggly,tension=0.2}{v2,m1}
\fmf{dbl_wiggly}{v4,m3}
\fmf{dbl_wiggly}{v3,m3}
\fmf{dbl_wiggly,tension=0.2}{v1,m1}
\fmf{dbl_wiggly,tension=1}{m1,m3}
\fmf{phantom}{i6,f2}
\fmf{dashes,foreground=red}{f2,f3}
\fmf{phantom}{f3,f4}
\fmf{phantom}{f4,ff5}
\fmf{phantom}{ff5,fff5}\fmf{phantom}{fff5,f5}
\fmf{phantom}{f5,f6}
\fmf{phantom}{f6,f8}
\fmf{phantom}{f8,o6}
\fmf{phantom}{i4,xf2}
\fmf{dashes,foreground=red}{xf2,xf3}
\fmf{phantom}{xf3,xf4}
\fmf{phantom}{xf4,xff5}
\fmf{phantom}{xff5,xfff5}\fmf{phantom}{xfff5,xf5}
\fmf{phantom}{xf5,xf6}
\fmf{phantom}{xf6,xf8}
\fmf{phantom}{xf8,o4}
\end{fmfgraph*}
\end{fmffile}
\end{gathered} \longrightarrow 
\begin{gathered}
    \begin{fmffile}{newother8}
    \begin{fmfgraph*}(100,100)
\fmfleftn{i}{9}
\fmfrightn{o}{9}
\fmf{phantom}{i1,vv1}
\fmf{phantom}{vv2,i9}
\fmf{phantom}{vv3,i5}
\fmf{phantom}{o5,vv4}
\fmf{phantom}{vv1,vv2}
\fmf{dbl_wiggly,tension=0.2}{vv1,v1}
\fmf{dbl_wiggly,tension=0.2}{v1,vv2}
\fmf{dbl_wiggly,tension=0.2}{v1,vv3}
\fmf{dbl_wiggly}{vv4,v3}
\fmf{dbl_wiggly,tension=1}{v1,v3}
\fmfv{decor.shape=circle,decor.filled=hatched, decor.size=3thick}{vv1}
\fmfv{decor.shape=circle,decor.filled=hatched, decor.size=3thick}{vv2}
\fmfv{decor.shape=circle,decor.filled=hatched, decor.size=3thick}{vv3}
\fmfv{decor.shape=circle,decor.filled=hatched, decor.size=3thick}{vv4}
\end{fmfgraph*}
\end{fmffile}
\end{gathered} $\\
$\begin{gathered}
    \begin{fmffile}{newother9}
    \begin{fmfgraph*}(100,100)
\fmfleftn{i}{9}
\fmfrightn{o}{9}
\fmf{plain,label.side=left}{i1,v1}
\fmf{plain,label.side=left}{v2,i9}
\fmf{plain,label.side=right}{v3,o9}
\fmf{plain,label.side=right}{o1,v3}
\fmf{plain,tension=0.3}{v1,vv5}
\fmf{plain,tension=0.3}{v2,vv5}
\fmf{plain,tension=0.3}{v5,vv5}
\fmf{plain,tension=0.3}{v5,vv5}
\fmf{phantom,tension=0}{v5,m1}
\fmf{dbl_wiggly,tension=0.2}{vv2,m1}
\fmf{dbl_wiggly,tension=0}{vv1,v5}
\fmf{dbl_wiggly,tension=0.2}{v2,vv2}
\fmf{dbl_wiggly}{v4,m3}
\fmf{dbl_wiggly}{v3,m3}
\fmf{dbl_wiggly,tension=0.2}{vv1,m1}
\fmf{dbl_wiggly,tension=0.2}{v1,vv1}
\fmf{dbl_wiggly,tension=1}{m1,m3}
\fmf{phantom}{i6,f2}
\fmf{dashes,foreground=red}{f2,f3}
\fmf{phantom}{f3,f4}
\fmf{phantom}{f4,ff5}
\fmf{phantom}{ff5,fff5}\fmf{phantom}{fff5,f5}
\fmf{phantom}{f5,f6}
\fmf{phantom}{f6,f8}
\fmf{phantom}{f8,o6}
\fmf{phantom}{i4,xf2}
\fmf{dashes,foreground=red}{xf2,xf3}
\fmf{phantom}{xf3,xf4}
\fmf{phantom}{xf4,xff5}
\fmf{phantom}{xff5,xfff5}\fmf{phantom}{xfff5,xf5}
\fmf{phantom}{xf5,xf6}
\fmf{phantom}{xf6,xf8}
\fmf{phantom}{xf8,o4}
\end{fmfgraph*}
\end{fmffile}
\end{gathered} \longrightarrow 
\begin{gathered}
    \begin{fmffile}{newother10}
    \begin{fmfgraph*}(100,100)
\fmfleftn{i}{9}
\fmfrightn{o}{9}
\fmf{phantom}{i1,vv1}
\fmf{phantom}{vv2,i9}
\fmf{phantom}{vv3,i5}
\fmf{phantom}{o5,vv4}
\fmf{phantom}{vv1,vv2}
\fmf{dbl_wiggly,tension=0.2}{vv1,vvv1}
\fmf{dbl_wiggly,tension=0.2}{v1,vvv1}
\fmf{dbl_wiggly,tension=0.2}{vvv2,vv2}
\fmf{dbl_wiggly,tension=0.2}{v1,vvv2}
\fmf{dbl_wiggly,tension=0}{vvv1,vv3}
\fmf{phantom,tension=0.2}{v1,vv3}
\fmf{dbl_wiggly}{vv4,v3}
\fmf{dbl_wiggly,tension=1}{v1,v3}
\fmfv{decor.shape=circle,decor.filled=hatched, decor.size=3thick}{vv1}
\fmfv{decor.shape=circle,decor.filled=hatched, decor.size=3thick}{vv2}
\fmfv{decor.shape=circle,decor.filled=hatched, decor.size=3thick}{vv3}
\fmfv{decor.shape=circle,decor.filled=hatched, decor.size=3thick}{vv4}
\end{fmfgraph*}
\end{fmffile}
\end{gathered} $
\end{tabular}
      \caption{Correspondence with world-line diagrams.}
      \label{fig:worldline}
    \end{figure}
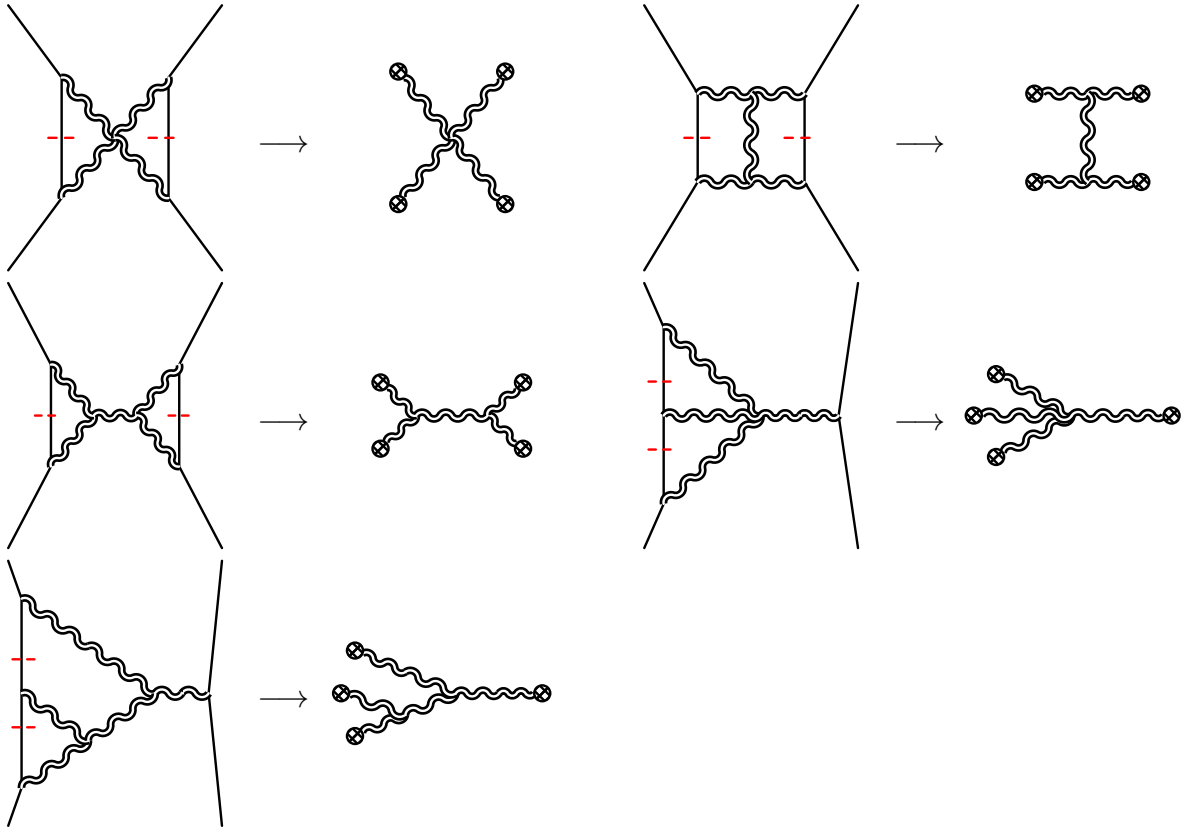

This graphical interpretation of the action of the delta functions also has another advantage that we wish to explain.
In ref. \cite{Bjerrum-Bohr:2018xdl} we showed how to relate the
second quantized field theory analysis to that of the world-line formalism. In the conservative
sector and when integrals are evaluated in the potential region only, this connection arises from extracting the residues of the time-components of the loop
momenta. The velocity cuts that we have illustrated in this two-loop calculation can be viewed as a covariant generalization of that reduction in dimensionality of the loop
integrals. For an $n$-loop amplitude the most important set of graph topologies in this respect is that of just $n$ massive propagators. When we apply velocity cuts on such graphs
all massive propagators will have been acted upon and what is left can be reinterpreted as external sources after a Fourier transform.
In Figure~\ref{fig:worldline} we draw those two-loop topologies which correspond to just two massive propagators. As each velocity cut 
removes one loop integration and effectively reduces 
the matter line to two external sources we denote them by blobs in
the same figure. The remaining three-dimensional integrals are what appear as a spatial Fourier transform in the world-line formalism. One of these integrations can be
taken to represent the parametrized world line and the remaining two integrations are analogous to the transformation into $b$-space.

  Recently, K\"alin, Liu, and Porto~\cite{Kalin:2020fhe} (see also refs.~\cite{Kalin:2020mvi,Mogull:2020sak,Jakobsen:2021smu})  
have shown how to derive the conservative part of the 
Post-Minkowskian scattering angle to the third Post-Minkowskian order considered here. The diagrams needed are precisely those listed in Figure~\ref{fig:worldline}. 
There are obviously many other diagrams contributing to the (conservative) classical part of the full amplitude but those that factorize do not contribute to 
physical observables and are
eliminated by either the Born subtractions, or the effective field theory matching, or the eikonal exponentiation, depending on which framework one prefers. Indeed, the
world-line formalism works at the level of the effective action and hence lives already in the exponent, without any need of subtractions. This still leaves certain
terms left over after the factorization and those diagrams (which still have un-cut massive propagators because they stem from topologies with, for $n$-loop graphs, 
with more than $n$ massive propagators) do not appear to have a simple diagrammatic interpretation in world-line language, although they are of course there. Graph topologies 
corresponding to radiation-reaction contributions also have more than $n$ massive propagators and will therefore also not immediately be amenable to this kind of world-line
interpretation. It is interesting to note that the recent calculation of the conservative part 
at fourth Post-Minkowskian order \cite{Bern:2021dqo} is organized along similar lines of the exponentiated action and it should therefore match quite directly to the
world-line formalism at that order.

  The correspondence between our present computation and the calculation at the same order in the world-line formalism appears to run deeper than this. Indeed, 
the basis of integrals used in ref.~\cite{Kalin:2020fhe} coincides with ours, through the following translation table: The set
$$
\{I_{11111}, I_{11211}, I_{01101},I_{11011}, I_{00211}, I_{00112}, I_{00111}\},
$$
of eq. (16) in ref.~\cite{Kalin:2020fhe} corresponds to
$$
\{\mathcal I^{0,0,1,1,1,1,1},\mathcal I^{0,0,1,1,1,1,2},\mathcal I^{0,0,0,1,0,1,1},
\mathcal I^{0,0,1,1,1,1,0},
\mathcal I^{0,0,0,0,1,1,2},
\mathcal I^{0,0,0,0,1,2,1},
\mathcal I^{0,0,0,0,1,1,1}\},
$$
in our notation of equation~(5.10) in~\cite{Bjerrum-Bohr:2021vuf}. This basis of master integrals suffices for the part of the conservative sector that is immediately translatable into those
velocity cut diagrams that have matches to standard world-line
diagrams. We note that $ \mathcal I^{0,0,0,1,0,1,1}=0$  is set to zero in the analysis
of~\cite{Kalin:2020fhe,Parra-Martinez:2020dzs}, and indeed it only
contributes to the radiation-reaction terms.
In ref.~\cite{Kalin:2020fhe} two more basis integrals were needed. They were denoted by
$$
\{M_{11,11100}^{(1,2)},M_{11,11100}^{(1,1)}\},
$$ 
in eq. (18) of ref.~\cite{Kalin:2020fhe} and  they correspond to our remaining two
$$
\{\mathcal I^{1,1,0,0,1,1,1},\mathcal I^{1,1,0,0,1,1,1} (\sigma=1)\} ,
$$
from ref.~\cite{Bjerrum-Bohr:2021vuf}. This demonstrates explicitly the one-to-one match of the two bases. It is as expected for the diagrams for the conservative sector
since we have seen the translation table between graph topologies. But as we have shown here, 
just the same basis of integrals also encapsulates all radiation-reaction parts. Since the basis of master integrals contains also the crucial terms from what is known as  the 
soft region of the integrals, it does suggest that radiation reaction also has a natural interpretation in world-line language. At higher orders in the Post-Minkowskian
expansion we expect the corresponding choice of a minimal basis of master integrals to be quite crucial in order to simplify calculations.

\subsection{The scattering angle}\label{sec:angle}

We can now evaluate the eikonal phase at third Post-Minkowskian
order.
The classical limit of the two-loop two-body scattering in~\ref{e:M2result}
takes the form needed for the exponentation of the classical third
Post-Minkowskian eikonal in~\ref{e:M2classical}

\begin{equation}
  1+i\sum_{L\geq0} {\widetilde {\mathcal M}}_L(\sigma,b)=
  (1+2i\Delta(\sigma,b)) \exp\left({2i\over\hbar}\sum_{L\geq0}\delta_L(\sigma,b)\right).
\end{equation}
Thanks to the factorized form of the two-loop classical contribution
in~\eqref{e:M2result} we have
 the first Post-Minkowskian eikonal phase 
\begin{equation}
\delta_0(\sigma,b)={\hbar\over2}\eqref{e:Mzero}=-\frac{G_N m_1
  m_2(2\sigma^2-1) }{2\epsilon \sqrt{\sigma^2-1}}(\pi b^2
e^{\gamma_E})^{\epsilon}+\mathcal O(\epsilon),
\end{equation}
the second Post-Minkowskian eikonal phase
\begin{equation}
\delta_1(\sigma,b)={\hbar\over2}\eqref{e:Mone}=\frac{3 \pi G_N^2 (m_1+m_2) m_1 m_2 (5\sigma^2-1)}{8 b \sqrt{\sigma^2-1}} (\pi b^2 e^{\gamma_E})^{2\epsilon},
\end{equation}
and the third Post-Minkowskian eikonal phase from the real part of the
third Post-Minkowskian amplitude
\begin{multline}\label{e:delta2}
\delta_2(\sigma,b)={\hbar\over2}\Re\eqref{e:M2classical}=\frac{G_N^3 m_1
  m_2 (\pi b^2e^{\gamma_E })^{3\epsilon}}{2b^2
  \sqrt{\sigma^2-1}}\Bigg(\frac{2s(12\sigma^4-10\sigma^2+1)}{\sigma^2-1}
\cr
-{4m_1m_2\sigma\over3}(25+14\sigma^2)+\frac{4 m_1 m_2(3+12
  \sigma^2-4\sigma^4)\arccosh(\sigma)}{\sqrt{\sigma^2-1}} \cr
+\frac{2 m_1 m_2(2\sigma^2-1)^2}{ \sqrt{\sigma^2-1}}
{1\over
  (4(\sigma^2-1))^{\epsilon}}\bigg(-\frac{11}{3}+\frac{d}{d\sigma}
\Big(\frac{(2\sigma^2-1)\arccosh(\sigma)}{\sqrt{\sigma^2-1}} \Big)
\bigg)\Bigg).
\end{multline}
And the leading quantum corrections at one-loop
\begin{equation}
  2\Delta_1=\widetilde{\mathcal M}_1^{\rm Qt.}(\sigma,b).
\end{equation}

\paragraph{The third Post-Minkowskian order scattering angle} is then obtained as
\begin{equation}
  \sin\left(\chi\over2\right)\Big|_{3PM}=-{\sqrt{s}\over m_1m_2\sqrt{\sigma^2-1}}{\partial
    \delta_2(\sigma,b)\over \partial b},
\end{equation}
giving
\begin{multline}
  \sin\left(\chi\over2\right)\Big|_{3PM}=  \frac{G_N^3 \sqrt{s}  }{b^3
 (\sigma^2-1)}\Bigg(\frac{3(2\sigma^2-1)(5\sigma^2-1)s}{2(\sigma^2-1)}\cr
+\frac{m_1^2+m_2^2}{2}(18\sigma^2-1)-{m_1
m_2\over3}\sigma(103+2\sigma^2)+\frac{4 m_1 m_2(3+12
  \sigma^2-4\sigma^4)\arccosh(\sigma)}{\sqrt{\sigma^2-1}} \cr
+\frac{2 m_1 m_2(2\sigma^2-1)^2}{\pi  \sqrt{\sigma^2-1}} {1\over
  (4(\sigma^2-1))^{\epsilon}}\bigg(-\frac{11}{3}+\frac{d}{d\sigma}
\Big(\frac{(2\sigma^2-1)\arccosh(\sigma)}{\sqrt{\sigma^2-1}} \Big)
\bigg)\Bigg)+\mathcal O(\epsilon).
\end{multline}
Using the definition of the angular momentum
\begin{equation}
  J={m_1m_2\sqrt{\sigma^2-1}\over\sqrt{s}}b\cos\left(\chi\over2\right),
\end{equation}
we obtain for the scattering angle at the first and second
Post-Minkowskian order 
\begin{align}
  \chi_{1PM}&={2G_Nm_1m_2(2\sigma^2-1)\over J\sqrt{\sigma^2-1}},\cr
\chi_{2PM}&={3\pi G_N^2m_1^2m_2^2(m_1+m_2)(5\sigma^2-1)\over 4J^2\sqrt{s}},
\end{align}
and at the third Post-Minkowksian order we obtain
\begin{equation}
  \chi_{3PM}= \widehat{\chi}_{3PM}+\chi_{3PM}^{\rm Rad.},
\end{equation}
with 
\begin{multline}
\widehat{  \chi}_{3PM}= \frac{2G_N^3 m_1^3 m_2^3 \left(64 \sigma
    ^6-120 \sigma ^4+60 \sigma ^2-5\right)}{3 J^3 \left(\sigma
    ^2-1\right)^{3\over2}}\cr
+{8G_N^3m_1^4m_2^4\sqrt{\sigma^2-1}\over 3J^3s}\Bigg(\sigma(-25-14\sigma^2)+\frac{3(3+12
  \sigma^2-4\sigma^4)\arccosh(\sigma)}{\sqrt{\sigma^2-1}}\Bigg),
\end{multline}
from the first and the  second line of~\eqref{e:delta2}, matching the
expressions in~\cite{Damour:2017zjx,Bern:2019nnu,Antonelli:2019ytb}.
And the radiation-reaction part from the third line of~\eqref{e:delta2}
\begin{equation}
  \chi_{3PM}^{\rm Rad.}= {4G_N^3m_1^4m_2^4(2\sigma^2-1)^2\over J^3s}
{1\over
  (4(\sigma^2-1))^{\epsilon}}\bigg(-\frac{11}{3}+\frac{d}{d\sigma}
\Big(\frac{(2\sigma^2-1)\arccosh(\sigma)}{\sqrt{\sigma^2-1}} \Big)
\bigg).
\end{equation}
This expression matches the results of~\cite{Damour:2020tta,DiVecchia:2021ndb,Herrmann:2021tct}.
\section{Conclusion}\label{sec:conclusion}

Using the integration method proposed in ref.~\cite{Bjerrum-Bohr:2021vuf}, we have evaluated the two-to-two massive scattering amplitude of two scalars
in Einstein gravity, keeping all terms that are needed to compute the classical scattering angle at third Post-Minkowskian order. The expansion is ordered
as a Laurent expansion in $\hbar$, but can also be broken into pieces corresponding to the conservative sector and additional pieces that have recently
been identified as radiation-reaction contributions. Our results have confirmed all aspects of the recent work in 
refs.~\cite{DiVecchia:2020ymx,Damour:2020tta,DiVecchia:2021ndb,DiVecchia:2021bdo}. 

An important simplification of  our calculation stems from the fact that the same small basis of master integrals that recently was used to solve the corresponding 
problem in maximal supergravity~\cite{Bjerrum-Bohr:2021vuf} can be used in Einstein gravity as well. Similarly, the reduction of dimensionality in the integrals 
due to a specific grouping of integrands works in Einstein gravity just as it does in maximal supergravity. This gives us the link to the world-line formalism through
what we have dubbed velocity cuts. We anticipate that these simplifications apply to higher orders in the Post-Minkowskian expansion as well.

\subsection*{Acknowledgements}
 We acknowledge interesting discussions with participants at the GGI workshop "Gravitational scattering, inspiral, and radiation", April-May, 2021. The research of P.V. has received funding from the ANR grant
``Amplitudes'' ANR-17- CE31-0001-01, and the ANR grant ``SMAGP''
ANR-20-CE40-0026-01 and is partially supported by Laboratory of Mirror
Symmetry NRU HSE, RF Government grant, ag. No 14.641.31.0001. P.V. is
grateful to the I.H.E.S. for the use of their computer resources. The
work of P.H.D. was supported in part by DFF grant 0135-00089A. The work of N.E.J.B.-B. was supported in part by the Carlsberg Foundation.

\appendix

\section{The five-point tree amplitude and the three-graviton cut}\label{sec:fiveamplitude}

To compute the cut integral, we need the five points amplitude where a
massive 
scalar emits three gravitons.  The result for these
contributions is obtained by computing the tree amplitudes using
Feynman diagrams, and was confirmed using the double copy  relations
(see~\cite{Bern:2019prr} for a review)
from the QCD amplitude of the emission of three gluons from a massive scalar~\cite{Forde:2005ue}.
This amplitude can be put in a simple form using the helicity formalism
(see~\cite{Mangano:1990by} for a review),
and subsequently considering the two independent helicity configurations~\cite{PlanteThesis}.

\paragraph{The singlet amplitude}
\begin{equation}\label{e:ggGGGhelppp}
i\mathcal M_0(p_1,p_1',l_1^+,l_2^+,l_3^+)  =-\frac{(8\pi G_N)^{3\over2} m_1^4}{\an[l_1,l_2]^2 \an[l_1,l_3]^2 \an[l_2,l_3]^2} \sum_{1 \leq i \ne j \ne k \leq 3} {(l_i\cdot l_j)(l_j\cdot l_k) tr_{+}[l_k,p_1,p_1',l_i]\over
    (p_1\cdot l_k)(p_1'\cdot l_i)},
\end{equation}
with $i\mathcal M_0(p_1,p_1',l_1^-,l_2^-,l_3^-)$ obtained by complex conjugation.
The singlet amplitude is the one given in~\cite{Bern:1998sv}. This amplitude vanishes when $m_1=0$.

We have defined
\begin{equation}
  tr_\pm(abcd)\equiv 2(a\cdot b c\cdot d-a\cdot c b\cdot d+a\cdot d b\cdot
  c)\pm 2i \epsilon^{\mu\nu\rho\sigma}a_\mu b_\nu c_\rho d_\sigma .
\end{equation}
\paragraph{The non-singlet amplitude}
 \begin{multline}
   \label{e:tggGGGhelmpp}
i\mathcal M_0(p_1,p_1',l_1^-,l_2^+,l_3^+)  ={ (2\pi G_N)^{3\over2}\over2}
 \Bigg(\sum_{2 \leq j \ne k \leq 3} \frac{
   \spab[l_1,p_1,l_j]\spab[l_1,p_1',l_j]^2\spab[l_1,p_1,l_k]^3}{\an[l_1,l_j]\an[l_1,l_k](l_1\cdot
   l_j)(l_1\cdot l_k)(p_1\cdot l_1)(p_1'\cdot l_j)} \\
 - \frac{
   \spab[l_1,p_1,l_2]^3\spab[l_1,p_1',l_3]^3}{\an[l_1,l_2]\an[l_1,l_3](l_1\cdot
   l_2)(l_1\cdot l_3)(p_1\cdot l_2)(p_1'\cdot l_3)}- \frac{2
   \sq[l_2,l_3] \spab[l_1,p_1,l_2]\spab[l_1,p_1,l_3]\langle l_1 | p_1
   | p_1' | l_1\rangle ^2}{\an[l_1,l_2]\an[l_1,l_3]\an[l_2,l_3](l_1\cdot l_2)(l_1\cdot l_3)(p_1\cdot l_1)}
 \\+\frac{2 \sq[l_2,l_3]^3 \langle l_1 | p_1 | p_1' | l_1\rangle^2}{\an[l_2,l_3](l_1\cdot l_2)(l_1\cdot l_3)t} \Bigg)+(p_1 \leftrightarrow -p_1'),
 \end{multline}
 with $i\mathcal M_0(p_1,p_1',l_1^+,l_2^-,l_3^-)$ obtained by complex conjugation.
The advantage of these expressions is that they keep track of the
symmetry regarding $p_1$ and $-p_1'$ exchange, and regarding the
internal momenta $l_1$, $l_2$ and $l_3$ satisfying $l_1+l_2+l_3+q=0$.

\paragraph{The three-particle cut}

The three-particle cut in~\eqref{e:3cut} is obtained by summing the
singlet and the non-singlet contributions
\begin{equation}
  \mathcal M_2^{\rm 3-cut}(\sigma,q^2)=  \mathcal M_2^{\rm
    3-cut}(\sigma,q^2)\Big|_{\rm singlet}+ \mathcal M_2^{\rm
    3-cut}(\sigma,q^2)\Big|_{\rm non-singlet},
\end{equation}
where
\begin{multline}
  \mathcal M_2^{\rm    3-cut}(\sigma,q^2)\Big|_{\rm singlet}=\int \frac{d^D l_1
  d^D  l_2}{(2\pi)^{2D}}\delta(l_1+l_2+l_3+q){i^3\over
  l_1^2l_2^2l_3^2}\cr
\times{1\over 3!}\Big(\sum_{\textrm{Perm}(l_1,l_2,l_3)}
 \mathcal    M_0(p_1,p_1',l_1^+,l_1^+,l_3^+) \mathcal
 M_0(p_2,p_2',l_1^-,l_1^-,l_3^-)\cr
+ {1\over 3!}\sum_{\textrm{Perm}(l_1,l_2,l_3)}\mathcal
    M_0(p_1,p_1',l_1^-,l_1^-,l_3^-) \mathcal
    M_0(p_2,p_2',l_1^+,l_1^+,l_3^+)\Big),
\end{multline}
and the non-singlet cut
\begin{multline}
  \mathcal M_2^{\rm
    3-cut}(\sigma,q^2)\Big|_{\rm non-singlet}=\int \frac{d^D l_1
  d^D  l_2}{(2\pi)^{2D}}\delta(l_1+l_2+l_3+q){i^3\over
  l_1^2l_2^2l_3^2}\cr
\times\Big({1\over 3!}\sum_{\textrm{Perm}(l_1,l_2,l_3)}
 \mathcal    M_0(p_1,p_1',l_1^-,l_1^+,l_3^+) \mathcal
 M_0(p_2,p_2',l_1^+,l_1^-,l_3^-)\cr
 +{1\over 3!}\sum_{\textrm{Perm}(l_1,l_2,l_3)}\mathcal
    M_0(p_1,p_1',l_1^+,l_1^-,l_3^-) \mathcal
    M_0(p_2,p_2',l_1^-,l_1^+,l_3^+)\Big).
\end{multline}

\section{Numerator factors}\label{sec:numerators}

In this appendix we list the numerator factors entering the expression
of the three-particle cut of section~\ref{sec:threecut}.
With the individual integral defined by the graphs we obtain the
following numerators from the tensorial reduction of the
three-particle cut:

\paragraph{The double-box numerators}
have to be expanded  up to order $(\hbar \underline{\vec q})^2$.

The numerator factor for the double-box integral in the left of
figure~\ref{fig:doublebox} in the $s$-channel
\begin{equation}
\mathcal N_{\dBox}^{(s)}=512\pi^3G_N^3(m_1^4+m_2^4-2(m_1^2+m_2^2)s+s^2)^3=2^{12}\pi^3G_N^3 m_1^6 m_2^6 (2\sigma^2-1)^3,
\end{equation}
the  numerator factor for the crossed double-box integral in the right of
figure~\ref{fig:doublebox} in the $s$-channel
\begin{equation}
\mathcal N_{\dBox}^{(cross,s)}=2^{13}\pi^3G_N^3 \left(96 m_1^6 m_2^6(2\sigma^2-1)^3+8 m_1^5 m_2^5 \sigma(2\sigma^2-1)^2(\hbar \underline{\vec q})^2 (l_2 \cdot l_3)+\mathcal O((\hbar \underline{\vec q})^4)\right).
\end{equation}
The numerator for the double-box integral in the $u$-channel is
given by
\begin{align}
  \mathcal N_{\dBox}^{(u)}&=512\pi^3G_N^3(m_1^4+m_2^4-2(m_1^2+m_2^2)u+u^2)^3\\
\nonumber  &=2^{12}\pi^3G_N^3\left(96 m_1^6
m_2^6(2\sigma^2-1)^3
-6 m_1^5 m_2^5\sigma (2\sigma^2-1)^2 (\hbar \underline{\vec q})^2+\mathcal O((\hbar \underline{\vec q})^4)\right),
\end{align}
and for the cross double-box integral in the $u$-channel
\begin{equation}
\mathcal N_{\dBox}^{(cross,u)}=2^{13}\pi^3G_N^3\left(m_1^6 m_2^6(2\sigma^2-1)^3-8 m_1^5 m_2^5\sigma(2\sigma^2-1)^2(\hbar \underline{\vec q})^2 (l_2 \cdot l_3+\frac{3}{4})+\mathcal O((\hbar \underline{\vec q})^4)\right).
\end{equation}

\paragraph{The box-triangle numerators} have to be expanded to the  order $|\hbar\underline{\vec{q}}|$.

The numerator factor for the non-planar box-triangle integral in the right of
figure~\ref{fig:boxtriangle} in the $s$-channel is $\mathcal
N_{\triangleboxright}^{(NP)}=\mathcal
N_{\triangleboxright}^{(NP,I)}+\mathcal
N_{\triangleboxright}^{(NP,II)}$ with 
\begin{multline}
\mathcal N_{\triangleboxright}^{(NP,I)}=1024\pi^3G_N^3\Big(\frac{m_1^6m_2^2(2\sigma^2-1)((p_2 \cdot l_1-p_2 \cdot l_2)^2+4m_2^2(1-4\sigma^2)(l_1 \cdot l_2))}{l_1 \cdot l_2}\\-\frac{2 m_1^5 m_2 \sigma (\hbar \underline{\vec q})(p_2 \cdot l_1-p_2 \cdot l_2)(2m_2^2(1-2\sigma^2)-(p_2 \cdot l_1-p_2 \cdot l_2)^2+8m_2^2(3\sigma^2-1)(l_1 \cdot l_2))}{l_1 \cdot l_2}\cr+\mathcal O((\hbar \underline{\vec q})^2)\Big),
\end{multline}
and
\begin{multline}
\mathcal N_{\triangleboxright}^{(NP,II)}=1024\pi^3G_N^3\Big(\frac{m_1^6m_2^2(2\sigma^2-1)((p_4 \cdot l_1-p_4 \cdot l_2)^2+4m_2^2(1-4\sigma^2)(l_1 \cdot l_2))}{l_1 \cdot l_2}\\-\frac{2 m_1^5 m_2 \sigma (\hbar \underline{\vec q})(p_2 \cdot l_1-p_2 \cdot l_2)(2m_2^2(1-2\sigma^2)-(p_2 \cdot l_1-p_2 \cdot l_2)^2+8m_2^2(3\sigma^2-1)(l_1 \cdot l_2))}{l_1 \cdot l_2}\cr+\mathcal O((\hbar \underline{\vec q})^2)\Big).
\end{multline}
The numerator factor for the planar box-triangle integral in the left of
figure~\ref{fig:boxtriangle} in the $s$-channel is $\mathcal
N_{\triangleboxright}^{(s)}=\mathcal
N_{\triangleboxright}^{(s,I)}+\mathcal
N_{\triangleboxright}^{(s,II)}$ with 
\begin{multline}
\mathcal N_{\triangleboxright}^{(s,I)}=1024\pi^3G_N^3\Big(\frac{24m_1^6m_2^2(2\sigma^2-1)((p_2 \cdot l_1-p_2 \cdot l_3)^2+4m_2^2(1-4\sigma^2)(l_1 \cdot l_3))}{l_1 \cdot l_3}\\-16 m_1^5 m_2^3 \sigma (\hbar \underline{\vec q})(p_2 \cdot l_1-p_2 \cdot l_3)(2\sigma^2-1)+\mathcal O((\hbar \underline{\vec q})^2)\Big),
\end{multline}
and
\begin{multline}
\mathcal N_{\triangleboxright}^{(s,II)}=1024\pi^3G_N^3\Big(\frac{m_1^6m_2^2(2\sigma^2-1)((p_4 \cdot l_1-p_4 \cdot l_3)^2+4m_2^2(1-4\sigma^2)(l_1 \cdot l_3))}{l_1 \cdot l_3}\\+16 m_1^5 m_2^3 \sigma (\hbar \underline{\vec q})(p_2 \cdot l_1-p_2 \cdot l_3)(2\sigma^2-1)+\mathcal O((\hbar \underline{\vec q})^2)\Big).
\end{multline}
The numerator factor for the planar box-triangle integral in the left of
figure~\ref{fig:boxtriangle} in the $u$-channel is $\mathcal
N_{\triangleboxright}^{(u)}=\mathcal
N_{\triangleboxright}^{(u,I)}+\mathcal
N_{\triangleboxright}^{(U,II)}$ with 
\begin{multline}
\mathcal N_{\triangleboxright}^{(u,I)}=1024\pi^3G_N^3\Big(\frac{m_1^6m_2^2(2\sigma^2-1)((p_4 \cdot l_1-p_4 \cdot l_3)^2+4m_2^2(1-4\sigma^2)(l_1 \cdot l_3))}{l_1 \cdot l_3}\\-16 m_1^5 m_2^3 \sigma (\hbar \underline{\vec q})(p_2 \cdot l_1-p_2 \cdot l_3)(2\sigma^2-1)+\mathcal O((\hbar \underline{\vec q})^2)\Big),
\end{multline}
and
\begin{multline}
\mathcal N_{\triangleboxright}^{(u,II)}=1024\pi^3G_N^3\Big(\frac{m_1^6m_2^2(2\sigma^2-1)((p_2 \cdot l_1-p_2 \cdot l_3)^2+4m_2^2(1-4\sigma^2)(l_1 \cdot l_3))}{l_1 \cdot l_3}\\+16 m_1^5 m_2^3 \sigma (\hbar \underline{\vec q})(p_2 \cdot l_1-p_2 \cdot l_3)(2\sigma^2-1)+\mathcal O((\hbar \underline{\vec q})^2)\Big),
\end{multline}
with similar expression for the numerators of the amplitude
$\mathcal M_2^{\triangleboxleft}(\sigma,b)$. 

\paragraph{The double-triangle numerator}
The numerator factor for the double-triangle diagram in
figure~\ref{fig:triangle} is given by 
\begin{multline}
\mathcal N_{\triangleright}=-1024\pi^3G_N^3\Big(\frac{m_1^6
  m_2^2(2\sigma^2-1)(u_q \cdot l_2-u_q \cdot l_3)^2}{l_2 \cdot l_3}+\frac{m_1^6
  m_2^2(2\sigma^2-1)(u_q \cdot l_1-u_q \cdot l_3)^2}{l_1 \cdot l_3}\cr
+\frac{m_1^6 m_2^2(2\sigma^2-1)(u_q \cdot l_1-u_q \cdot l_2)^2}{(l_1 \cdot l_2)}- m_1^6 m_2^2(-5+34 \sigma^2)+\mathcal O((\hbar \underline{\vec q}))\Big).
\end{multline}

\paragraph{The $H$ diagram numerator}
The numerator factor for the $H$ diagram in
figure~\ref{fig:Hdiagram} is given by 
\begin{multline}
\mathcal N_{H}={128\pi^3G_N^3\over3}\Big(-48(-4 m_1^2 m_2^4
((l_2+l_3)^2 - (l_1+l_3)^2 + 4 \sigma^2) (\bar{p}_{1} \cdot l_2)^2 \cr
- 
  8 m_2^4 (\bar{p}_{1} \cdot l_2)^4 + 16 m_1^3 m_2^3 \sigma (\bar{p}_{1} \cdot l_2)(\bar{p}_{2} \cdot l_1)\\ + 
  m_1^4 \Big(m_2^4 (-1 - 2 (l_2+l_3)^2 (1 + (l_2+l_3)^2) - 2
  (l_1+l_3)^2 (1 +(l_1+l_3)^2) \cr
  + 4 \sigma^2 + 
  4 ((l_2+l_3)^2 + (l_2+l_3)^4 + (l_1+l_3)^2 
  - 2 (l_2+l_3)^2(l_1+l_3)^2 + (l_1+l_3)^4\Big) \sigma^2 \cr
  - 
        4 \sigma^4) + 4 m_2^2 ((l_2+l_3)^2 - (l_1+l_3)^2 - 4 \sigma^2) (\bar{p}_{2} \cdot l_1)^2 - 
     8 (\bar{p}_{2} \cdot l_1)^4))(\hbar \underline{\vec q})^4+\mathcal O((\hbar \underline{\vec q})^5)\Big).
\end{multline}

\paragraph{The box-bubble diagram numerator}
The numerator factor for the box diagram in
figure~\ref{fig:box} is given by
\begin{equation}
\mathcal N_{\Box}=\frac{2048\pi^3G_N^3m_1^4m_2^4(2\sigma^2-1)(1+522 \sigma^2)}{15}.
\end{equation}

\section{The one-loop two-body amplitude}\label{sec:oneloop}
The one-loop two-body scattering is done using the two-particle cuts
following the computation in~\cite{Bjerrum-Bohr:2013bxa}
\begin{equation}
 {\mathcal M}^2_{\rm 2-cut}(\sigma,q^2)=
\qquad  \begin{gathered}
 \begin{fmffile}{2cut}
 \begin{fmfgraph*}(100,100)
\fmfstraight
\fmfleftn{i}{2}
\fmfrightn{o}{2}
\fmftop{t}
\fmfbottom{b}
\fmfrpolyn{smooth,filled=30,label=\textrm{tree}}{el}{8}
\fmfrpolyn{smooth,filled=30,label=\textrm{tree}}{er}{8}
\fmf{fermion,label=$p_1$,label.side=left,tension=2}{i1,el1}
\fmf{fermion,label=$p_1'$,label.side=left,tension=2}{el3,i2}
\fmf{fermion,label=$p_2'$,tension=2}{er5,o2}
\fmf{fermion,label=$p_2$,label.side=right,tension=2}{o1,er7}
\fmf{dbl_wiggly,tension=.1}{el5,er3}
\fmf{dbl_wiggly,tension=.1}{el7,er1}
\fmf{dashes,for=red}{b,t}
\end{fmfgraph*}
\end{fmffile}
\end{gathered} \qquad .
\end{equation}
We reduce the tensorial integrals using {\tt
  LiteRed}~\cite{Lee:2013mka}, we find that the amplitude decomposes as
\begin{equation}\label{e:M1total}
 \mathcal M_1(\sigma,q^2)=\mathcal M_1^{\Box}+\mathcal
 M_1^{\triangleright}+\mathcal M_1^{\triangleleft}+\mathcal M_1^{\circ}.
\end{equation}

\paragraph{The box contribution} is given by
\begin{equation}
    \mathcal M_1^{\Box}(\sigma,q^2)=256 \pi^2 m_1^4 m_2^4 G_N^2 (2\sigma^2-1)^2
  (I_s+I_u)-1024 \pi^2 m_1^3 m_2^3 G_N^2 (2\sigma^2-1)|\hbar\underline
  q|^2 I_u ,
\end{equation}
with the scalar one-loop box and cross-box
integrals 
\begin{equation}
  I_s=  \begin{gathered}
    \begin{fmffile}{box}
    \begin{fmfgraph*}(100,100)
\fmfstraight
\fmfleftn{i}{2}
\fmfrightn{o}{2}
\fmf{fermion,label=$p_1$,label.side=left}{i1,v1}
\fmf{fermion,label=$p_1'$,label.side=left}{v2,i2}
\fmf{fermion,label=$p_2'$,label.side=right}{v3,o2}
\fmf{fermion,label=$p_2$,label.side=right}{o1,v4}
\fmf{fermion}{v1,v2}
\fmf{fermion}{v4,v3}
\fmf{dbl_wiggly}{v4,v1}
\fmf{dbl_wiggly}{v2,v3}
\end{fmfgraph*}
\end{fmffile}
\end{gathered}, \qquad
I_u=
\begin{gathered}
    \begin{fmffile}{crossbox}
    \begin{fmfgraph*}(100,100)
\fmfstraight
\fmfleftn{i}{2}
\fmfrightn{o}{2}
\fmf{fermion,label=$p_1$,label.side=left}{i1,v1}
\fmf{fermion,label=$p_1'$,label.side=left}{v2,i2}
\fmf{fermion,label=$p_2'$,label.side=right}{v3,o2}
\fmf{fermion,label=$p_2$,label.side=right}{o1,v4}
\fmf{fermion,tension=0.01}{v1,v2}
\fmf{fermion,tension=0.01}{v4,v3}
\fmf{dbl_wiggly}{v4,v2}
\fmf{dbl_wiggly}{v1,v3}
\fmf{phantom,tension=0}{v1,v2}
\fmf{phantom,tension=0}{v4,v3}
\end{fmfgraph*}
\end{fmffile}
\end{gathered} \, .
\end{equation}
This box part has the following Laurent expansion in $\hbar$ up to the
first quantum correction
\begin{equation}
  \mathcal M_1^{\Box}(\sigma,q^2)= {1\over |\underline
    q|^{2\epsilon}}\left( \mathcal M_1^{\Box(-2)}+\mathcal M_1^{\Box(-1)}+\mathcal M_1^{\Box(0)}+\mathcal O(\hbar) \right),
\end{equation}
with
\begin{equation}
  \mathcal M_1^{\Box(-2)}=-\frac{8 \pi m_1^3 m_2^3 G_N^2
                           (2\sigma^2-1)^2 (4\pi)^{\epsilon}
                           \Gamma(1+\epsilon)
                           \Gamma(-\epsilon)^2}{|\hbar\underline
                           {q}|^{2}\Gamma(-2\epsilon)
                           \sqrt{\sigma^2-1}},
                       \end{equation}
\begin{equation}                       
 \label{e:Boxclassical}\mathcal M_1^{\Box(-1)}=8\pi^2 m_1^2m_2^2 G_N^2 (2\sigma^2-1)^2
                          \frac{i(4\pi)^\epsilon (m_1+m_2)}{
                          |\hbar\underline {q}|
                          (\sigma^2-1)}\frac{\Gamma(\frac{1}{2}-\epsilon)^2
                          \Gamma(\frac{1}{2}+\epsilon)}{\pi^{3\over
                            2}\Gamma(-2\epsilon)},
                      \end{equation}
\begin{align}                      
  \mathcal M_1^{\Box(0)}&=      
4 m_1^2 m_2^2 G_N^2 (2\sigma^2-1)^2
                           \frac{(4\pi)^{\epsilon}\Gamma(1+\epsilon)\Gamma(-\epsilon)^2}{
                                                      (\sigma^2-1)^{\frac{3}{2}}\Gamma(-2\epsilon)}\cr
&         \times                  \left(\frac{\epsilon s \pi }{2 m_1 m_2
                           }+i(1+2\epsilon)(\sigma
                           \arccosh(\sigma)-\sqrt{\sigma^2-1})\right)\cr
  \label{e:Boxquantum}                   &-32 m_1^2 m_2^2 G_N^2 (2\sigma^2-1) \frac{i(4\pi)^\epsilon \arccosh(\sigma)}{ \sqrt{\sigma^2-1}}\frac{\Gamma(-\epsilon)^2 \Gamma(1+\epsilon)}{\Gamma(-2\epsilon)} |\hbar\underline{q}|^2\,.
\end{align}
At the  leading order in $\epsilon$ the first quantum correction from
the box integral reads
\begin{multline}
\mathcal M_1^{\Box(0)}=-\frac{8G_N^2m_1^2 m_2^2(2\sigma^2-1)(4\pi
  e^{-\gamma_E})^{\epsilon}}{\epsilon |\vec{q}|^{2\epsilon}
  (\sigma^2-1)^{\frac{3}{2}}} \cr
\times\left(\frac{(2\sigma^2-1)\epsilon s \pi }{2 m_1 m_2 }+i((7-6\sigma^2)\arccosh(\sigma)-(2\sigma^2-1)\sqrt{\sigma^2-1})\right).
\end{multline}

\paragraph{The triangle contribution:}
The right-triangle graph
\begin{equation}
 \mathcal M_1^{\triangleright}(\sigma,q^2)=\begin{gathered}
    \begin{fmffile}{righttriangle}
    \begin{fmfgraph*}(100,100)
\fmfstraight
\fmfleftn{i}{2}
\fmfrightn{o}{2}
\fmf{fermion,label=$p_1$,label.side=left}{i1,v1}
\fmf{fermion,label=$p_1'$,label.side=left}{v2,i2}
\fmf{fermion,label=$p_2'$,label.side=right}{v4,o2}
\fmf{fermion,label=$p_2$,label.side=right}{o1,v4}
\fmf{fermion,tension=0.01}{v1,v2}
\fmf{dbl_wiggly}{v1,v4}
\fmf{dbl_wiggly}{v2,v4}
\fmf{phantom,tension=0}{v1,v2}
\end{fmfgraph*}
\end{fmffile}
\end{gathered}
\end{equation}
has the following numerator factor
\begin{equation}
\mathcal N_{\triangleright}=64 \pi^2 m_1^4 G_N^2 \left(4m_2^2
  (4\sigma^2-1)+\frac{8
    (p_2\cdot l_1)^2}{|\hbar\underline{q}|^2}\right)-256 \pi^2 m_1^3 G^2
(m_1+8 m_2 \sigma)p_2\cdot l_1 +\mathcal O(|\hbar\underline{q}|^2),
\end{equation}
which  after tensorial reduction leads to
\begin{multline}
    \mathcal M_1^{\triangleright}(\sigma,q^2)= 24 i \pi^2 G_N^2 m_1^4m_2^2(5\sigma^2-1)\int \frac{d^D l_1}{(2\pi)^{D-1}}
  \frac{\delta(p_1\cdot l_1)}{l_1^2 (l_1+q)^2} \cr-64 \pi^2 G_N^2 m_1^2 m_2^2
  (22 \sigma^2-1) \int \frac{d^D l_1}{(2\pi)^D} \frac{1}{l_1^2
    (l_1+q)^2}.
\end{multline}

The Laurent expansion in $\hbar$ up to the first quantum correction reads
\begin{equation}
      \mathcal M_1^{\triangleright}(\sigma,q^2)={1\over
        |q|^{2\epsilon}}\left(   \mathcal M_1^{\triangleright(-1)}+
        \mathcal M_1^{\triangleright(0)}+\mathcal O(\hbar)\right),
\end{equation}
with the classical and first quantum contributions
\begin{align}
  \label{e:triangleclassical}
  \mathcal M_1^{\triangleright(-1)}&=\frac{3 i \pi^2 G_N^2 m_1^3m_2^2(5\sigma^2-1) (4\pi e^{-\gamma_E})^{\epsilon}}{ |\hbar\underline{q}| },\\
\label{e:trianglequantum} \mathcal M_1^{\triangleright(0)}&=- \frac{4 i G_N^2 m_1^2 m_2^2 ( 22 \sigma^2-1)i (4\pi e^{-\gamma_E})^{\epsilon}}{\epsilon } \,           ,  
\end{align}
with an equivalent expression for the left-triangle graph $
\mathcal M_1^{\triangleleft}(\sigma,q^2)$ with exchanging  $m_1$
and $m_2$.
\paragraph{The bubble contribution}
\begin{equation}
 \mathcal M_1^{\circ}(\sigma,q^2)=
  \begin{gathered}
    \begin{fmffile}{bubble}
    \begin{fmfgraph*}(100,100)
\fmfstraight
\fmfleftn{i}{2}
\fmfrightn{o}{2}
\fmf{fermion,label=$p_1$,label.side=left}{i1,v1}
\fmf{fermion,label=$p_1'$,label.side=left}{v1,i2}
\fmf{fermion,label=$p_2'$}{v4,o2}
\fmf{fermion,label=$p_2$,label.side=right}{o1,v4}
\fmf{fermion,tension=0.01}{v1,v2}
\fmf{dbl_wiggly,right}{v1,v4}
\fmf{dbl_wiggly,left}{v1,v4}
\fmf{phantom,tension=0}{v1,v2}
\end{fmfgraph*}
\end{fmffile}
\end{gathered} 
\end{equation}
has the numerator factor
\begin{multline}
\mathcal N_{\circ}=64 \pi^2 G_N^2 \Big(-4 m_1^2 m_2^2 + 40 m_1^2 m_2^2
\sigma^2 + 24 m_2^2 (p_1\cdot l_1)^2\cr - 
 64 m_1 m_2 \sigma p_1\cdot l_1 p_2\cdot l_1 + 24 m_1^2 (p_2\cdot l_1)^2 + 
 16 (p_1\cdot l_1)^2 (p_2\cdot l_1)^2\Big) +\mathcal O(|\hbar \underline{q}|),
\end{multline}
which after tensorial reduction leads to
\begin{equation}
  \mathcal M_1^{\circ}(\sigma,q^2)=  \frac{64 \pi^2 G_N^2 m_1^2 m_2^2 (1 + 522 \sigma^2)}{15} \int \frac{d^D l_1}{(2\pi)^D} \frac{1}{l_1^2 (l_1+q)^2}.
\end{equation}
This starts contributing from the first quantum correction 
\begin{equation}\label{e:bubble}
  \mathcal M_1^{\circ(0)}(\sigma,q^2)=\frac{4 i (4\pi e^{-\gamma_E})^{\epsilon}G_N^2 m_1^2 m_2^2 (1 + 522 \sigma^2)}{15  \epsilon |\underline{q}|^{2\epsilon}}.
\end{equation}

\paragraph{The one-loop  amplitude } in~\eqref{e:M1total}, and reads  including  the first quantum correction 
\begin{equation}\label{e:M1expand}
  \mathcal M_1(\sigma,q^2)={1\over
        |q|^{2\epsilon}}\left( \mathcal M_1^{(-2)}+  \mathcal
        M_1^{(-1)}+  \mathcal M_1^{(0)}+\mathcal O(\hbar)\right),
\end{equation}
with
\begin{align}
  \mathcal M_1^{(-2)}&=\mathcal M_1^{\Box(-2)},\cr
                       \mathcal M_1^{(-1)}&=\mathcal
                                            M_1^{\Box(-1)}+\mathcal
                                            M_1^{\triangleright(-1)}+\mathcal
                                            M_1^{\triangleleft(-1)},\cr
 \mathcal M_1^{(0)}&=\mathcal M_1^{\Box(0)}+\mathcal
                                            M_1^{\triangleright(0)}+\mathcal
                                            M_1^{\triangleleft(0)}+\mathcal
                     M_1^{\circ(0)}.
\end{align}
\section{Master integrals}\label{sec:masters}
In this appendix we recall the
expressions for the master integrals computed
in~\cite{Bjerrum-Bohr:2021vuf}.

With the following definition for the master integral
\begin{multline}\label{e:Imasterdef}
\!\!\!\!\!\!\mathcal I^{n_1,n_2,n_3,n_4,n_5,n_6,n_7}\equiv\int
\frac{d^{D-1}l_1d^{D-1}l_2}{(2\pi)^{2D-2}} \frac{1}{(k \cdot
  l_1)^{n_1} (k \cdot l_2)^{n_2}(l_1^2)^{n_3}((u_q+l_2)^2)^{n_4}
  ((l_1-u_q)^2)^{n_5} }\cr
\times \frac{1}{ (l_2^2)^{n_6}((l_1+l_2)^2-2(\sigma-1) k \cdot l_1 k \cdot l_2)^{n_7}},
\end{multline}
where we have  defined $k^2\equiv u_q^2\equiv-1$ and $k\cdot u_q\equiv
0$, we collect  the results obtained in~section~5 of~\cite{Bjerrum-Bohr:2021vuf}.

\begin{equation}
  \mathcal I_1(\sigma)\equiv2\epsilon^4 \mathcal I^{0,0,1,1,1,1,0}
                        = \epsilon^4 \frac{\Gamma(\frac{1}{2}+\epsilon)^2 \Gamma(\frac{1}{2}-\epsilon)^4}{2(4\pi)^{3-2\epsilon} \Gamma(1-2\epsilon)^2},
\end{equation}
and
\begin{align}
 \mathcal I_2(\sigma)&\equiv2\epsilon^4 \sqrt{\sigma^2-1}\mathcal I^{0,0,0,0,1,1,1}\cr
                        &= \Bigg(-\frac{2\epsilon \Gamma(\frac{1}{2}-\epsilon)^3\Gamma(2\epsilon)}{(4\pi)^{D-1}\Gamma(\frac{3}{2}-3\epsilon)} (\sigma^2-1)^{3\epsilon} \int_{1}^{\sigma}
  {dt\over (t^2-1)^{\frac{1}{2}+3\epsilon}
  }\cr
  &+2b_5(\sigma^2-1)^{3\epsilon}  \int_{1}^{\sigma}
 {dt_2\over  (t_2^2-1)^{\frac{1}{2}+3\epsilon}}\int_{1}^{t_2}{dt_1\over (t_1^2-1)^{\frac{1}{2}+\epsilon}} \Bigg)\epsilon^3  +\mathcal O(\epsilon^4),
\end{align}
and
\begin{align}
\mathcal I_3(\sigma)&=2\epsilon^3 \sqrt{\sigma^2-1}\mathcal I^{0,0,0,0,1,1,2}\cr
                        &={-b_5 \epsilon\over (\sigma^2-1)^{\epsilon}}
  -\Big(-4{b_5\over (\sigma^2-1)^{\epsilon}} \int_{1}^{\sigma}
 {dt_2 \over  (t_2^2-1)^{\frac{1}{2}-\epsilon} }\int_{1}^{t_2}
 {dt_1\over (t_1^2-1)^{\frac{1}{2}+\epsilon}}\cr
 &+\frac{4\epsilon
   \Gamma(\frac{1}{2}-\epsilon)^3\Gamma(2\epsilon)}{(4\pi)^{D-1}\Gamma(\frac{3}{2}-3\epsilon)}
   {1\over (\sigma^2-1)^{\epsilon} } \int_{1}^{\sigma} {dt\over (t^2-1)^{\frac{1}{2}-\epsilon}}\Big)\epsilon^3+\mathcal O(\epsilon^4),
\end{align}
and
\begin{align}
 \mathcal I_4(\sigma)&\equiv4\epsilon^2 (\sigma^2-1)\mathcal I^{-1,-1,0,0,1,1,3}+\epsilon^2(1+2\epsilon) \sigma \mathcal I^{0,0,0,0,1,1,2}\cr
                        &=\left(-\frac{2\epsilon \Gamma(\frac{1}{2}-\epsilon)^3\Gamma(2\epsilon)}{(4\pi)^{D-1}\Gamma(\frac{3}{2}-3\epsilon)}+2 b_5 \int_{1}^{\sigma} {dt\over (t^2-1)^{\frac{1}{2}+\epsilon}}\right)\epsilon^2+\mathcal O(\epsilon^4),
\end{align}
and
\begin{align}
 \mathcal I_5(\sigma)&\equiv\frac{2\epsilon^2(4\epsilon-1)(2\epsilon-1)}{\sqrt{\sigma^2-1}}\mathcal I^{0,0,0,1,0,1,1}\cr
                        &=-\frac{i\epsilon  (4\pi)^{2\epsilon} }{32\pi^2  (\sigma^2-1)^{\epsilon}}\frac{(-1)^\epsilon \Gamma(1-2\epsilon)^2 
\Gamma(1+2\epsilon)^2 \Gamma(1-\epsilon)}{\Gamma(1-4\epsilon)\Gamma(1+\epsilon)},
\end{align}
with $(-1)^\epsilon=e^{i\pi \epsilon}=1+i \pi \epsilon +\mathcal
O(\epsilon^2)$,
and
\begin{align}
  \mathcal I_6(\sigma)&\equiv2\epsilon^4 \sqrt{\sigma^2-1}\mathcal I^{0,0,1,1,1,1,1}\\
\nonumber                        &=\frac{(4\pi e^{- \gamma_E})^{2\epsilon}
  \epsilon^3}{8\pi^3}
\arcsinh\left(\sqrt{\frac{\sigma-1}{2}}\right)\!\!\left(\pi\!+\!2i\left(-1\over  4(\sigma^2-1)\right)^{\epsilon}  \arcsinh\left(\sqrt{\frac{\sigma-1}{2}}\right)\!\!\right)+\mathcal O(\epsilon^4),
\end{align}
and
\begin{align}
 \mathcal I_{7}(\sigma)&\equiv8\epsilon^4(\sigma^2-1)\mathcal I^{-1,-1,1,1,1,1,1}+4\epsilon^4 \sigma \mathcal I^{0,0,1,1,1,1,0},\cr
                        &=0+\mathcal O(\epsilon^3),
\end{align}
and
\begin{equation}
 \mathcal I_8(\sigma)\equiv-\epsilon^3 \mathcal I^{0,1,0,0,1,1,2}
                        =\pm \frac{i \sqrt{\pi} \epsilon^3 \Gamma(-\epsilon)\Gamma(-\frac{1}{2}-2\epsilon)\Gamma(\frac{3}{2}+2\epsilon)\Gamma(\frac{1}{2}-\epsilon)\Gamma(-\frac{1}{2}-\epsilon)}{(4\pi)^{3-2\epsilon}\Gamma(-2\epsilon)\Gamma(-\frac{1}{2}-3\epsilon)},
                      \end{equation}
with the $+$ sign for the $+i\varepsilon$    prescription and $-$ for
the $+i\varepsilon$ of the linear denominator. We refer to section~5.5.4 of~\cite{Bjerrum-Bohr:2021vuf} for details.
And, finally,
\begin{align}
  \mathcal I_9(\sigma)&\equiv\epsilon^4 \mathcal I^{1,1,0,0,1,1,1},\\
                        &=b_9 \epsilon^2-{b_5 \epsilon^2 \over (\sigma^2-1)^\epsilon}\Bigg( \arccosh(\sigma)
                          -\epsilon\Big(\arccosh(\sigma)^2\cr
\nonumber  & \hskip5cm+\Li_2\left(2-2\sigma(\sigma+\sqrt{\sigma^2-1})\right)
\Big) +\mathcal O(\epsilon)\Bigg)\,.
\end{align}
with
\begin{equation}
  \label{eq:3}
  b_5=-\frac{i  (4\pi)^{2\epsilon} }{32\pi^2  }\frac{(-1)^\epsilon \Gamma(1-2\epsilon)^2 
\Gamma(1+2\epsilon)^2 \Gamma(1-\epsilon)}{\Gamma(1-4\epsilon)\Gamma(1+\epsilon)},
\end{equation}
and $b_9$ depends on the $i\varepsilon$ prescription of the linear
denominators, with the result (we refer to section~5.5.5 of~\cite{Bjerrum-Bohr:2021vuf} for details)
\begin{equation}
b_9^{++}=b_9^{--}=-2b_9^{+-}.
\end{equation}
with
\begin{equation}\label{e:b9}
b_9^{+-}=b_9^{-+}=-\frac{\epsilon^2}{6} \frac{\Gamma(-\epsilon)^3\Gamma(1+2\epsilon)}{(4\pi)^{2-2\epsilon} \Gamma(-3\epsilon)}=-\frac{(4\pi e^{-\gamma_E})^{2\epsilon}}{32 \pi^2}+\mathcal O(\epsilon^2).
\end{equation}


\end{document}